\documentstyle[12pt,epsf]{article}

\newcommand{\n}{\hspace*{-2.5mm}}
\newcommand{\li}{\mathop{{\mbox{Li}}_4}\nolimits}
\newcommand{\agt}{\,\rlap{\lower 3.5 pt \hbox{$\mathchar \sim$}} \raise 1pt
 \hbox {$>$}\,}

\catcode`@=11
\newcount\@tempcntc
\def\@citex[#1]#2{\if@filesw\immediate\write\@auxout{\string\citation{#2}}\fi
  \@tempcnta\z@\@tempcntb\m@ne\def\@citea{}\@cite{\@for\@citeb:=#2\do
    {\@ifundefined
       {b@\@citeb}{\@citeo\@tempcntb\m@ne\@citea\def\@citea{,}{\bf ?}\@warning
       {Citation `\@citeb' on page \thepage \space undefined}}%
    {\setbox\z@\hbox{\global\@tempcntc0\csname b@\@citeb\endcsname\relax}%
     \ifnum\@tempcntc=\z@ \@citeo\@tempcntb\m@ne
       \@citea\def\@citea{,}\hbox{\csname b@\@citeb\endcsname}%
     \else
      \advance\@tempcntb\@ne
      \ifnum\@tempcntb=\@tempcntc
      \else\advance\@tempcntb\m@ne\@citeo
      \@tempcnta\@tempcntc\@tempcntb\@tempcntc\fi\fi}}\@citeo}{#1}}
\def\@citeo{\ifnum\@tempcnta>\@tempcntb\else\@citea\def\@citea{,}%
  \ifnum\@tempcnta=\@tempcntb\the\@tempcnta\else
   {\advance\@tempcnta\@ne\ifnum\@tempcnta=\@tempcntb \else \def\@citea{--}\fi
    \advance\@tempcnta\m@ne\the\@tempcnta\@citea\the\@tempcntb}\fi\fi}
\catcode`@=12

\begin{document}

\title{\vskip-3cm{\baselineskip14pt
\centerline{\normalsize\hfill MPI/PhT/97--041}
\centerline{\normalsize\hfill hep--ph/9708255}
\centerline{\normalsize\hfill July 1997}
}
\vskip1.5cm
Decoupling Relations to ${\cal O}(\alpha_s^3)$
and their Connection to Low-Energy Theorems
}
\author{{\sc K.G. Chetyrkin}\thanks{Permanent address:
Institute for Nuclear Research, Russian Academy of Sciences,
60th October Anniversary Prospect 7a, Moscow 117312, Russia.}
{\sc, B.A. Kniehl, and M. Steinhauser}\\
Max-Planck-Institut f\"ur Physik (Werner-Heisenberg-Institut),\\
F\"ohringer Ring 6, 80805 Munich, Germany}
\date{}
\maketitle

\thispagestyle{empty}

\begin{abstract}
If quantum chromodynamics (QCD) is renormalized by minimal subtraction (MS),
at higher orders, the strong coupling constant $\alpha_s$ and the quark
masses $m_q$ exhibit discontinuities at the flavour thresholds, which are
controlled by so-called decoupling constants, $\zeta_g$ and $\zeta_m$,
respectively.
Adopting the modified MS ($\overline{\mbox{MS}}$) scheme, we derive simple
formulae which reduce the calculation of $\zeta_g$ and $\zeta_m$ to the
solution of vacuum integrals.
This allows us to evaluate $\zeta_g$ and $\zeta_m$ through three loops.
We also establish low-energy theorems, valid to all orders, which relate the
effective couplings of the Higgs boson to gluons and light quarks, due to the
virtual presence of a heavy quark $h$, to the logarithmic derivatives w.r.t.\
$m_h$ of $\zeta_g$ and $\zeta_m$, respectively.  
Fully exploiting present knowledge of the anomalous dimensions of $\alpha_s$ 
and $m_q$, we thus calculate these effective couplings through four loops.
Finally, we perform a similar analysis for the coupling of the Higgs boson to
photons.
\medskip

\end{abstract}

\newpage

\section{Introduction}

As is well known, in renormalization schemes based on the method of minimal
subtraction (MS) \cite{tHo73}, including the modified MS
($\overline{\mbox{MS}}$) scheme \cite{BarBurDukMut78}, which is routinely used
in quantum chromodynamics (QCD), the  Appelquist-Carazzone decoupling theorem
\cite{AppCar75} does not hold true in its na\"\i ve sense.
Let us consider QCD with $n_l=n_f-1$ light quarks flavours $q$ and one heavy
flavour $h$.
Then, the contributions of $h$ to the Green functions of the gluons and
light quarks expressed in terms of the renormalized parameters of the full
theory do not exhibit the expected $1/m_h$ suppression, where $m_h$ is the
mass of $h$.
The reason for this is that the $\beta$ and $\gamma_m$ functions
governing the running of the strong coupling constant $\alpha_s(\mu)$
and the light-quark masses $m_q(\mu)$ with the renormalization scale $\mu$
do not depend on any mass.

The standard procedure to circumvent this problem is to render decoupling
explicit by using the language of effective field theory, i.e.\ $h$ is 
integrated out.
Specifically, one constructs an effective $n_l$-flavour theory by requiring
consistency with the full $n_f$-flavour theory at the heavy-quark threshold
$\mu^{(n_f)}={\cal O}(m_h)$ \cite{Wei80,BerWet82}.
This leads to nontrivial matching conditions between the couplings and 
light-quark masses of the two theories.
Although, $\alpha_s^{(n_l)}(m_h)=\alpha_s^{(n_f)}(m_h)$ and
$m_q^{(n_l)}(m_h)=m_q^{(n_f)}(m_h)$ at leading (tree-level) and
next-to-leading (one-loop) order, these identities do not generally hold at
higher orders in the $\overline{\rm MS}$ scheme.
Starting at next-to-next-to-leading (two-loop) order, they are broken by
finite corrections, of ${\cal O}(\alpha_s^2)$, as was noticed in the 
pioneering works of Refs.~\cite{Wei80,BerWet82}.
The relations between the couplings and light-quark masses of the full and
effective theories are called decoupling relations; the proportionality
constants that appear in these relations are denoted decoupling constants,
$\zeta_g$ and $\zeta_m$.
In this paper, $\zeta_g$ and $\zeta_m$ are computed through
next-to-next-to-next-to-leading (three-loop) order, ${\cal O}(\alpha_s^3)$.
They have to be applied whenever a flavour threshold is to be crossed.
If the $\mu$ evolutions of $\alpha_s^{(n_f)}(\mu)$ and $m_q^{(n_f)}(\mu)$ are
to be performed at $N$ loops, then consistency requires that the decoupling 
relations be implemented at $N-1$ loops.
Then, the residual $\mu$ dependence of physical observables will be of
the $(N+1)$-loop order.
If our new results are combined with the recently evaluated four-loop
coefficients of the $\beta$ \cite{LarRitVer971} and $\gamma_m$
\cite{Che97LarRitVer972} functions, then it is possible to consistently
describe QCD-related observables through ${\cal O}(\alpha_s^4)$
\cite{CheKniSte97als}.

The dominant subprocess for the production of the standard-model (SM) Higgs
boson at the CERN Large Hadron Collider (LHC) will be the one via gluon
fusion.
Therefore, an important ingredient for the Higgs-boson search will be the
effective coupling of the Higgs boson to gluons, usually called $C_1$.
A standard approach to study this coupling as well as the Higgs-boson 
coupling to photons has been to use low-energy theorems (LET's) \cite{let}.
At the two-loop level, $C_1$ has been known since long
\cite{InaKubOka83,DjoSpiZer91}.
Recently, the three-loop correction to $C_1$ has been obtained through a 
direct diagrammatic calculation \cite{CheKniSte97}.
In Ref.~\cite{CheKniSte97}, also a LET which allows for
the computation of $C_1$ from the knowledge of $\zeta_g$ has been used.
In a similar way, the effective couplings of the Higgs boson to light quarks
may be treated as well.
The corresponding coefficient function, which is usually denoted by $C_2$, has
a similar connection to $\zeta_m$.
In this paper, we establish the relationships between $C_1,C_2$ and
$\zeta_g,\zeta_m$ to all orders by formulating appropriate LET's.
With their help, we compute $C_1$ and $C_2$ through four loops, i.e.\
${\cal O}(\alpha_s^4)$.

The outline of the paper is as follows.
In Section~\ref{seccon}, we derive simple formulae which allow us to 
determine the decoupling constants $\zeta_g$ and $\zeta_m$ for $\alpha_s(\mu)$
and $m_q(\mu)$, respectively, by simply evaluating vacuum integrals.
With the help of these formulae, in Section~\ref{secmat}, we calculate
$\zeta_g$ and $\zeta_m$ up to the three-loop order.
In Section~\ref{seclet}, we derive LET's, valid to all orders, for the
coefficient functions $C_1$ and $C_2$, which comprise the virtual top-quark
effects on the interactions of the Higgs boson with gluons and quarks,
respectively.
Exploiting these LET's, we compute $C_1$ and $C_2$ through four loops.
In Section~\ref{secgam}, we extend our analysis to also include the
interaction of the Higgs boson with photons.
In Section~\ref{secdis}, we explore the phenomenological significance of our 
results and present our conclusions.
In Appendix~\ref{appnc}, we display the $N_c$ dependence of our key results,
adopting the general gauge group SU($N_c$).
In Appendix~\ref{appz2z3}, we list the decoupling constants for the quark and
gluon fields in the covariant gauge.

\section{\label{seccon}Formalism}

Let us, in a first step, fix the notation and present the framework for our
calculation.
Throughout this paper, we work in the $\overline{\mbox{MS}}$ renormalization
scheme \cite{BarBurDukMut78}.
In the main part of the paper, we concentrate on the QCD gauge group SU(3), 
i.e.\ we put $N_c=3$.
The $\beta$ and $\gamma_m$ functions of QCD are defined through
\begin{eqnarray}
\frac{\mu^2d}{d\mu^2}\,\frac{\alpha_s^{(n_f)}}{\pi}&\n=\n&
\beta^{(n_f)}\left(\alpha_s^{(n_f)}\right)=
-\sum_{N=1}^\infty\beta_{N-1}^{(n_f)}
\left(\frac{\alpha_s^{(n_f)}}{\pi}\right)^{N+1},
\label{eqrgea}\\ 
\frac{1}{m_q^{(n_f)}}\,\frac{\mu^2d}{d\mu^2}m_q^{(n_f)}&\n=\n&
\gamma_m^{(n_f)}\left(\alpha_s^{(n_f)}\right) =
-\sum_{N=1}^\infty\gamma_{m,N-1}^{(n_f)}
\left(\frac{\alpha_s^{(n_f)}}{\pi}\right)^N,
\label{eqrgem}
\end{eqnarray}
where $N$ denotes the number of loops.
Recently, the four-loop coefficients $\beta_3^{(n_f)}$ and
$\gamma_{m,3}^{(n_f)}$ have become available
\cite{LarRitVer971,Che97LarRitVer972}.
According to present knowledge, we have
\begin{eqnarray}
\beta_0^{(n_f)} &\n=\n&\frac{1}{4}\left[ 11 - \frac{2}{3} n_f\right],
\nonumber\\
\beta_1^{(n_f)} &\n=\n&\frac{1}{16}\left[ 102 - \frac{38}{3} n_f\right],
\nonumber \\
\beta_2^{(n_f)} &\n=\n&\frac{1}{64}\left[\frac{2857}{2} - \frac{5033}{18} n_f
 + \frac{325}{54} n_f^2\right],
\nonumber \\
\beta_3^{(n_f)} &\n=\n&\frac{1}{256}\left[  \frac{149753}{6} + 3564 \zeta(3) 
        + \left(- \frac{1078361}{162} - \frac{6508}{27} \zeta(3) \right) n_f
\right.\nonumber \\
&\n\n&{}+ \left( \frac{50065}{162} + \frac{6472}{81} \zeta(3) \right) n_f^2
+\left.  \frac{1093}{729}  n_f^3\right],
\nonumber \\
\gamma_{m,0}^{(n_f)} &\n=\n& 1,
\nonumber\\
\gamma_{m,1}^{(n_f)} &\n=\n& \frac{1}{16}\left[ \frac{202}{3}
                 - \frac{20}{9} n_f \right],       
\nonumber \\
\gamma_{m,2}^{(n_f)} &\n=\n& \frac{1}{64} \left[1249+\left( - \frac{2216}{27} 
          - \frac{160}{3}\zeta(3) \right)n_f 
               - \frac{140}{81} n_f^2 \right],
\nonumber \\
\gamma_{m,3}^{(n_f)} &\n=\n& \frac{1}{256} \left[ 
       \frac{4603055}{162} + \frac{135680}{27}\zeta(3) - 8800\zeta(5)
 +\left(- \frac{91723}{27} - \frac{34192}{9}\zeta(3) 
    + 880\zeta(4) 
 \right.\right.
\nonumber \\
&\n\n&{}+ \left.\left.
 \frac{18400}{9}\zeta(5) \right) n_f
 +\left( \frac{5242}{243} + \frac{800}{9}\zeta(3) 
    - \frac{160}{3}\zeta(4) \right) n_f^2
 +\left(- \frac{332}{243} + \frac{64}{27}\zeta(3) \right) n_f^3 \right],
\nonumber\\
&\n\n&
\end{eqnarray}
where $\zeta$ is Riemann's zeta function, with values $\zeta(2)=\pi^2/6$,
$\zeta(3)\approx1.202\,057$, $\zeta(4)=\pi^4/90$, and
$\zeta(5)\approx1.036\,928$.

The relations between the bare and renormalized quantities read
\begin{eqnarray}
g_s^0&\n=\n&\mu^{2\varepsilon}Z_gg_s,\qquad
m_q^0=Z_mm_q,\qquad
\xi^0-1=Z_3(\xi-1),
\nonumber\\
\psi_q^0&\n=\n&\sqrt{Z_2}\psi_q,\qquad
G_\mu^{0,a}=\sqrt{Z_3}G_\mu^a,\qquad
c^{0,a}=\sqrt{\tilde{Z}_3}c^a,
\label{eqren}
\end{eqnarray}
where $g_s=\sqrt{4\pi\alpha_s}$ is the QCD gauge coupling,
$\mu$ is the renormalization scale,
$D=4-2\varepsilon$ is the dimensionality of space time,
$\psi_q$ is a quark field with mass $m_q$,
$G_\mu^a$ is the gluon field, and
$c^a$ is the Faddeev-Popov-ghost field.
For simplicity, we do not display the colour indices of the quark fields.
The gauge parameter, $\xi$, is defined through the gluon propagator in lowest
order,
\begin{equation}
\frac{i}{q^2+i\epsilon}\left(-g^{\mu\nu}+\xi\frac{q^\mu q^\nu}{q^2}\right).
\label{eqcov}
\end{equation}
The index `0' marks bare quantities.
We should mention that relations~(\ref{eqren}) hold true both in the full
$n_f$-flavour and effective $n_l$-flavour theories.
Only the renormalization constants $Z_g$ and $Z_m$ are relevant for our
purposes.
They are known through order ${\cal O}(\alpha_s^4)$ 
\cite{LarRitVer971,Che97LarRitVer972}.
The other relations are only listed in order to fix the notation.

In a similar way, we can write down the relations between the parameters and
fields of the full and effective theories.
From the technical point of view, it is simpler to first consider these
relations for bare quantities and to construct those for the renormalized ones
afterwards.
The bare decoupling constants are defined through the following equations:
\begin{eqnarray}
g_s^{0\prime}&\n=\n&\zeta_g^0 g_s^0,\qquad
m_q^{0\prime}=\zeta_m^0m_q^0,\qquad
\xi^{0\prime}-1=\zeta_3^0(\xi^0-1),
\nonumber\\
\psi_q^{0\prime}&\n=\n&\sqrt{\zeta_2^0}\psi_q^0,\qquad
G_\mu^{0\prime,a}=\sqrt{\zeta_3^0}G_\mu^{0,a},\qquad
c^{0\prime,a}=\sqrt{\tilde\zeta_3^0}c^{0,a},
\label{eqdec}
\end{eqnarray}  
where the primes mark the quantities of the effective $n_l$-flavour theory.

Combining Eqs.~(\ref{eqren}) and (\ref{eqdec}), we obtain relations between
the renormalized coupling constants and masses of the full and effective
theories, viz
\begin{eqnarray}
\alpha_s^\prime(\mu)&\n=\n& 
\left(\frac{Z_g}{Z_g^\prime}\zeta_g^0\right)^2\alpha_s(\mu)
=\zeta_g^2\alpha_s(\mu),
\label{eqdecg}\\
m_q^\prime(\mu)&\n=\n&
\frac{Z_m}{Z_m^\prime}\zeta_m^0m_q(\mu) 
=\zeta_mm_q(\mu).
\label{eqdecm}
\end{eqnarray}
Here, it is understood that the right-hand sides are expressed in terms of the
parameters of the full theory.
The unknown quantities in these decoupling relations are $\zeta_g^0$ and
$\zeta_m^0$.

In a next step, let us introduce the effective Lagrangian, ${\cal L}^\prime$,
which depends on the decoupling constants $\zeta_i^0$ and on the bare
parameters and light fields of the full theory.
It is obvious from gauge invariance that the most general form of
${\cal L}^\prime$ is simply the one of usual QCD, ${\cal L}^{\rm QCD}$,
retaining, however, only the light degrees of freedom.
Specifically, the definition of ${\cal L}^\prime$ reads
\begin{equation}
{\cal L}^\prime\left(g_s^0,m_q^0,\xi^0;\psi_q^0,G_\mu^{0,a},c^{0,a};\zeta_i^0
\right)
={\cal L}^{\rm QCD}\left(g_s^{0\prime},m_q^{0\prime},\xi^{0\prime};
\psi_q^{0\prime},G_\mu^{0\prime,a},c^{0\prime,a}\right),
\end{equation}
where $q$ represents the $n_l$ light quark flavours and $\zeta_i^0$
collectively denotes all bare decoupling constants of Eq.~(\ref{eqdec}).
Exploiting the circumstance that the result for some $n$-particle Green
function of light fields obtained from ${\cal L}^{\rm QCD}$ in the full theory
agrees, up to terms suppressed by inverse powers of the heavy-quark mass, with
the corresponding evaluation from ${\cal L}^\prime$ in the effective theory,
we may derive relations which allow us to determine the decoupling constants
$\zeta_i^0$.

As an example, let us consider the massless-quark propagator.
Up to terms of ${\cal O}(1/m_h)$, we have
\begin{eqnarray}
-\frac{1}{\not\!p\left[1+\Sigma_V^0(p^2)\right]}
&\n=\n&
i\int dx\,e^{ip\cdot x}\left\langle T\psi_q^0(x)\bar\psi_q^0(0)\right\rangle
\nonumber\\
&\n=\n&
\frac{i}{\zeta_2^0}
\int dx\,e^{ip\cdot x}\left\langle T\psi_q^{0\prime}(x)\bar\psi_q^{0\prime}(0)
\right\rangle
\nonumber\\
&\n=\n&
-\frac{1}{\zeta_2^0}\,
\frac{1}{\not\!p\left[1+\Sigma_V^{0\prime}(p^2)\right]},
\end{eqnarray}
where the subscript `V' reminds us that the QCD self-energy of a massless
quark only consists of a vector part.
Note that $\Sigma_V^{0\prime}(p^2)$ only contains light degrees of freedom,
whereas $\Sigma_V^0(p^2)$ also receives virtual contributions from the heavy
quark $h$.
As we are interested in the limit $m_h\to\infty$, we may nullify the external
momentum $p$, which entails an enormous technical simplification because then
only tadpole integrals have to be considered \cite{GorLar87}.
Within dimensional regularization, one also has $\Sigma_V^{0\prime}(0)=0$.
Thus, we obtain
\begin{equation}
\zeta_2^0=1+\Sigma_V^{0h}(0),
\label{eqdeccon0}
\end{equation}
where the superscript {\it h} indicates that only the hard part of the
respective quantities needs to be computed, i.e.\ only the diagrams involving
the heavy quark contribute.

In a similar fashion, it is possible to derive formulae for the
decoupling constants $\zeta_m^0$, $\zeta_3^0$, and $\tilde\zeta_3^0$ as well.
These read
\begin{eqnarray}
\zeta_m^0&\n=\n&\frac{1-\Sigma_S^{0h}(0)}{1+\Sigma_V^{0h}(0)},
\label{eqdeccon1}\\
\zeta_3^0&\n=\n&1+\Pi_G^{0h}(0),\\
\tilde\zeta_3^0&\n=\n&1+\Pi_c^{0h}(0),
\label{eqdeccon2}
\end{eqnarray}
where $\Sigma_V(p^2)$ and $\Sigma_S(p^2)$ are the vector and scalar components
of the light-quark self-energy, defined through
$\Sigma(p)={\not\!p}\Sigma_V(p^2)+m_q\Sigma_S(p^2)$, and $\Pi_G(p^2)$ and
$\Pi_c(p^2)$ are the gluon and ghost vacuum polarizations, respectively.
Specifically, $\Pi_G(p^2)$ and $\Pi_c(p^2)$ are related to the gluon and ghost
propagators through
\begin{eqnarray}
\delta^{ab}\left\{\frac{g^{\mu\nu}}{p^2\left[1+\Pi_G^0(p^2)\right]}
+\mbox{terms proportional to $p^\mu p^\nu$}\right\}
&\n=\n&i\int dx\,e^{ip\cdot x}
\left\langle TG^{0,a\mu}(x)G^{0,b\nu}(0)\right\rangle,
\nonumber\\
-\frac{\delta^{ab}}{p^2\left[1+\Pi_c^0(p^2)\right]}
&\n=\n&i\int dx\,e^{ip\cdot x}
\left\langle Tc^{0,a}(x)\bar{c}^{0,b}(0)\right\rangle,
\end{eqnarray}
respectively.
Finally, we extract an expression for $\zeta_g^0$ from the relationship
between the bare $G\bar cc$-vertex form factors of the full and effective
theories,
\begin{equation}
g_s^{0\prime}\left[1+\Gamma_{G\bar cc}^{0\prime}(p,k)\right]
=\frac{g_s^0}{\tilde\zeta_3^0\sqrt{\zeta_3^0}}
\left[1+\Gamma_{G\bar cc}^0(p,k)\right],
\end{equation}
by nullifying the external four-momenta $p$ and $k$.
Here, $\Gamma_{G\bar cc}^0(p,k)$ is defined through the
one-particle-irreducible (1PI) part of the amputated $G\bar cc$ Green function
as
\begin{eqnarray}
\lefteqn{p^\mu g_s^0\left\{-if^{abc}\left[1+\Gamma_{G\bar cc}^0(p,k)\right]
+\mbox{other colour structures}\right\}}
\nonumber\\
&\n=\n&i^2\int dxdy\,e^{i(p\cdot x+k\cdot y)}
\left\langle Tc^{0,a}(x)\bar c^{0,b}(0)G^{0,c\mu}(y)\right\rangle^{\rm 1PI},
\end{eqnarray}
where $p$ and $k$ are the outgoing four-momenta of $c$ and $G$, respectively,
and $f^{abc}$ are the structure constants of the QCD gauge group.
We thus obtain
\begin{equation}
\zeta_g^0=\frac{\tilde\zeta_1^0}{\tilde\zeta_3^0\sqrt{\zeta_3^0}},
\label{zetag0}
\end{equation}
where
\begin{equation}
\tilde\zeta_1^0=1+\Gamma_{G\bar cc}^{0h}(0,0).
\end{equation}

In contrast to the renormalization constants $Z_i$ in Eq.~(\ref{eqren}),
the decoupling constants $\zeta_i^0$ also receive contributions from the
finite parts of the loop integrals.
Thus, at ${\cal O}(\alpha_s^3)$, we are led to evaluate three-loop tadpole
integrals also retaining their finite parts.

\section{\label{secmat}Decoupling relations}

\begin{figure}[t]
\epsfxsize=16cm
\epsffile[74 624 569 725]{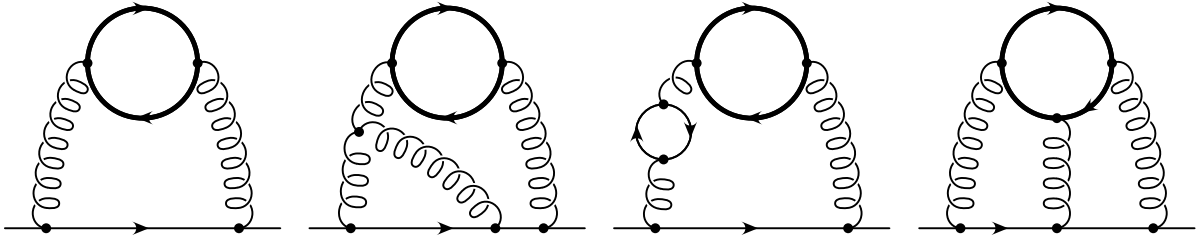}
\smallskip
\caption{Typical three-loop diagrams pertinent to $\Sigma_V^{0h}(0)$ and
$\Sigma_S^{0h}(0)$.
Solid, bold-faced, and loopy lines represent massless quarks $q$, heavy quarks
$h$, and gluons $G$, respectively.}
\label{f1}
\end{figure}

In this section, we compute the bare decoupling constants $\zeta_g^0$ and
$\zeta_m^0$ and combine them with the well-known ${\cal O}(\alpha_s^3)$
results for $Z_g$ and $Z_m$ to obtain the finite quantities $\zeta_g$ and
$\zeta_m$.
As may be seen from Eq.~(\ref{eqdeccon1}), the computation of $\zeta_m^0$
requires the knowledge of the hard part of the light-quark propagator.
There are no one-loop diagrams contributing to $\Sigma_V^{0h}(0)$ and
$\Sigma_S^{0h}(0)$.
At two loops, there is just one diagram.
Even at three loops, there is only a moderate number of diagrams, namely 25.
Typical specimen are depicted in Fig.~\ref{f1}.
Actually, through three loops, Eq.~(\ref{eqdeccon1}) simplifies to
$\zeta_m^0=1-\Sigma_V^{0h}(0)-\Sigma_S^{0h}(0)$.
It should be noted that the vector and scalar parts separately still depend on
the QCD gauge parameter $\xi$, but $\xi$ drops out in their sum, which is a
useful check for our calculation.
We use the program QGRAF \cite{Nog93} to generate the relevant diagrams.
The computation is then performed with the help of the program MATAD
\cite{Ste96}, which is based on the technology developed in Ref.~\cite{Bro92}
and written in FORM \cite{FORM}.
Using Eq.~(\ref{eqdecm}), we finally obtain
\begin{eqnarray}
\zeta_m&\n=\n&1
+\left(\frac{\alpha_s^{(n_f)}(\mu)}{\pi}\right)^2
\left(\frac{89}{432} 
-\frac{5}{36}\ln\frac{\mu^2}{m_h^2}
+\frac{1}{12}\ln^2\frac{\mu^2}{m_h^2}\right)
+\left(\frac{\alpha_s^{(n_f)}(\mu)}{\pi}\right)^3
\left[\frac{2951}{2916} 
-\frac{407}{864}\zeta(3)\right.
\nonumber\\
&\n\n&{}+\left.
\frac{5}{4}\zeta(4)
-\frac{1}{36}B_4
+\left(-\frac{311}{2592}
-\frac{5}{6}\zeta(3)\right)\ln\frac{\mu^2}{m_h^2}
+\frac{175}{432}\ln^2\frac{\mu^2}{m_h^2}
+\frac{29}{216}\ln^3\frac{\mu^2}{m_h^2}\right.
\nonumber\\
&\n\n&{}+\left.
n_l\left(
\frac{1327}{11664}
-\frac{2}{27}\zeta(3)
-\frac{53}{432}\ln\frac{\mu^2}{m_h^2}
-\frac{1}{108}\ln^3\frac{\mu^2}{m_h^2}\right)\right]
\nonumber\\
&\n\approx\n&1
+0.2060\left(\frac{\alpha_s^{(n_f)}(\mu_h)}{\pi}\right)^2
+\left(1.8476+0.0247\,n_l\right)
\left(\frac{\alpha_s^{(n_f)}(\mu_h)}{\pi}\right)^3,
\label{eqzetam}
\end{eqnarray}
where \cite{Bro92}
\begin{eqnarray}
B_4&\n=\n&16\li\left({1\over2}\right)-{13\over2}\zeta(4)-4\zeta(2)\ln^22
+{2\over3}\ln^42
\nonumber\\
&\n\approx\n&-1.762\,800,
\end{eqnarray}
with $\li{}$ being the quadrilogarithm, is a constant typical for three-loop
tadpole diagrams, $n_l=n_f-1$ is the number of light-quark flavours $q$, and
$m_h=m_h^{(n_f)}(\mu)$ is the $\overline{\mbox{MS}}$ mass of the heavy quark 
$h$.
For the numerical evaluation in the last line of Eq.~(\ref{eqzetam}),
we have chosen $\mu=\mu_h$, where $\mu_h=m_h^{(n_f)}(\mu_h)$.
The ${\cal O}(\alpha_s^2)$ term of Eq.~(\ref{eqzetam}) agrees with
Ref.~\cite{BerWet82}; the ${\cal O}(\alpha_s^3)$ term represents a new result.
The generalization of Eq.~(\ref{eqzetam}) appropriate for the gauge group
SU($N_c$) is listed in Appendix~\ref{appnc}.

For the convenience of those readers who prefer to fix the matching scale 
$\mu$ in units of the pole mass, $M_h$, we substitute in Eq.~(\ref{eqzetam}) 
the well-known relation between $m_h^{(n_f)}(\mu)$ and $M_h$
\cite{GraBroGraSch90} to obtain
\begin{eqnarray}
\zeta_m^{\rm OS}&\n=\n&1
+\left(\frac{\alpha_s^{(n_f)}(\mu)}{\pi}\right)^2
\left(\frac{89}{432} 
     -\frac{5}{36}\ln\frac{\mu^2}{M_h^2}
     +\frac{1}{12}\ln^2\frac{\mu^2}{M_h^2}
\right)
+\left(\frac{\alpha_s^{(n_f)}(\mu)}{\pi}\right)^3
\left[
\frac{1871}{2916} 
- \frac{407}{864}\zeta(3)
\right.
\nonumber\\
&\n\n&{}+
\left.
 \frac{5}{4}\zeta(4)
- \frac{1}{36}B_4
+\left(\frac{121}{2592}
- \frac{5}{6}\zeta(3)\right)\ln\frac{\mu^2}{M_h^2}
+ \frac{319}{432}\ln^2\frac{\mu^2}{M_h^2}
+ \frac{29}{216}\ln^3\frac{\mu^2}{M_h^2}
\right.
\nonumber\\
&\n\n&{}+
\left.
n_l\left(
\frac{1327}{11664}
- \frac{2}{27}\zeta(3)
- \frac{53}{432}\ln\frac{\mu^2}{M_h^2}
- \frac{1}{108}\ln^3\frac{\mu^2}{M_h^2}
\right)
\right]
\nonumber\\
&\n\approx\n&1
+0.2060\left(\frac{\alpha_s^{(n_f)}(M_h)}{\pi}\right)^2
+\left(1.4773+0.0247\,n_l\right)
\left(\frac{\alpha_s^{(n_f)}(M_h)}{\pi}\right)^3.
\label{eqzetamos}
\end{eqnarray}

\begin{figure}[t]
\epsfxsize=16cm
\epsffile[136 634 470 722]{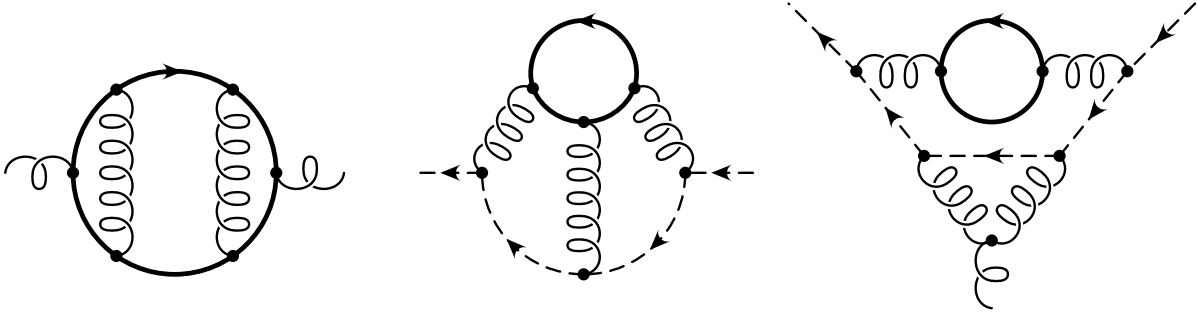}
\smallskip
\caption{Typical three-loop diagrams pertinent to $\Pi_G^{0h}(0)$,
$\Pi_c^{0h}(0)$, and $\Gamma_{G\bar cc}^{0h}(0,0)$.
Bold-faced, loopy, and dashed lines represent heavy quarks $h$, gluons $G$, 
and Faddeev-Popov ghosts $c$, respectively.}
\label{f2}
\end{figure}

According to Eq.~(\ref{zetag0}), three ingredients enter the calculation of
$\zeta_g^0$, namely the hard parts of the gluon and ghost propagators and the
gluon-ghost vertex correction. 
At one loop, only one diagram contributes, namely the diagram where the gluon
splits into a virtual pair of heavy quarks.
At two loops, three diagrams contribute to $\Pi_G^{0h}(0)$ and one to
$\Pi_c^{0h}(0)$.
In the case of $\Gamma_{G\bar cc}^{0h}(0,0)$, there are five diagrams, which,
however, add up to zero.
To this order, the three contributions are still separately independent of the
gauge parameter $\xi$, so that the $\xi$ independence of their combination
does not provide a meaningful check for our calculation.
The situation changes at ${\cal O}(\alpha_s^3)$, where all three parts
separately depend on $\xi$ and only their proper combination is $\xi$
independent as is required for a physical quantity.
At this order, the numbers of diagrams pertinent to $\Pi_G^{0h}(0)$,
$\Pi_c^{0h}(0)$, and $\Gamma_{G\bar cc}^{0h}(0,0)$ are 189, 25, and 228,
respectively.
Typical representatives are shown in Fig.~\ref{f2}.
The complexity of the problem at hand necessitates the use of powerful
analytic technology \cite{Nog93,Ste96,Bro92,FORM} to organize the calculation.
Inserting into Eq.~(\ref{eqdecg}) the result for $\zeta_g^0$ thus obtained
finally leads to the following answer in the $\overline{\mbox{MS}}$ scheme:
\begin{eqnarray}
\zeta_g^2&\n=\n&1
+\frac{\alpha_s^{(n_f)}(\mu)}{\pi}
\left(
-\frac{1}{6}\ln\frac{\mu^2}{m_h^2}
\right)
+\left(\frac{\alpha_s^{(n_f)}(\mu)}{\pi}\right)^2
\left(
\frac{11}{72} 
-\frac{11}{24}\ln\frac{\mu^2}{m_h^2}
+\frac{1}{36}\ln^2\frac{\mu^2}{m_h^2}
\right)
\nonumber\\
&\n\n&{}+\left(\frac{\alpha_s^{(n_f)}(\mu)}{\pi}\right)^3
\left[
\frac{564731}{124416} 
-\frac{82043}{27648}\zeta(3)
-\frac{955}{576}\ln\frac{\mu^2}{m_h^2}
+\frac{53}{576}\ln^2\frac{\mu^2}{m_h^2}
-\frac{1}{216}\ln^3\frac{\mu^2}{m_h^2} 
\right.
\nonumber\\
&\n\n&{}+
\left.
n_l\left(
-\frac{2633}{31104}
+\frac{67}{576}\ln\frac{\mu^2}{m_h^2} 
-\frac{1}{36}\ln^2\frac{\mu^2}{m_h^2}
\right)
\right]
\nonumber\\
&\n\approx\n&1
+0.1528\left(\frac{\alpha_s^{(n_f)}(\mu_h)}{\pi}\right)^2
+\left(0.9721-0.0847\,n_l\right)
\left(\frac{\alpha_s^{(n_f)}(\mu_h)}{\pi}\right)^3.
\label{eqzetag}
\end{eqnarray}
The ${\cal O}(\alpha_s^3)$ term in Eq.~(\ref{eqzetag}) has recently been
published \cite{CheKniSte97als}.
Leaving aside this term, Eq.~(\ref{eqzetag}) agrees with
Ref.~\cite{LarRitVer95}, while the constant term in ${\cal O}(\alpha_s^2)$
slightly differs from the result published in Ref.~\cite{BerWet82}.
In the meantime, the authors of Ref.~\cite{BerWet82} have revised \cite{Ber97}
their original analysis and have found agreement with Ref.~\cite{LarRitVer95}.
The SU($N_c$) version of Eq.~(\ref{eqzetag}) may be found in
Appendix~\ref{appnc}.
Introducing the pole mass $M_h$ leads to
\begin{eqnarray}
\left(\zeta_g^{\rm OS}\right)^2&\n=\n&1
+\frac{\alpha_s^{(n_f)}(\mu)}{\pi}
\left(
-\frac{1}{6}\ln\frac{\mu^2}{M_h^2}
\right)
+\left(\frac{\alpha_s^{(n_f)}(\mu)}{\pi}\right)^2
\left(
-\frac{7}{24} 
-\frac{19}{24}\ln\frac{\mu^2}{M_h^2}
+\frac{1}{36}\ln^2\frac{\mu^2}{M_h^2}
\right)
\nonumber\\
&\n\n&{}+\left(\frac{\alpha_s^{(n_f)}(\mu)}{\pi}\right)^3
\left[
-\frac{58933}{124416}
-\frac{2}{3}\zeta(2)\left(1+\frac{1}{3}\ln2\right)
-\frac{80507}{27648}\zeta(3)
-\frac{8521}{1728}\ln\frac{\mu^2}{M_h^2}
\right.\nonumber\\
&\n\n&{}-\left.
\frac{131}{576}\ln^2\frac{\mu^2}{M_h^2}
-\frac{1}{216}\ln^3\frac{\mu^2}{M_h^2} 
+n_l\left(
\frac{2479}{31104}
+\frac{\zeta(2)}{9}
+\frac{409}{1728}\ln\frac{\mu^2}{M_h^2} 
\right)
\right]
\nonumber\\
&\n\approx\n&1
-0.2917\left(\frac{\alpha_s^{(n_f)}(M_h)}{\pi}\right)^2
+\left(-5.3239+0.2625\,n_l\right)
\left(\frac{\alpha_s^{(n_f)}(M_h)}{\pi}\right)^3.
\label{eqzetagos}
\end{eqnarray}
Notice that the ${\cal O}(\alpha_s^3)$ terms of $\zeta_m$ and $\zeta_g$ depend
on the number $n_l$ of light (massless) quark flavours.
However, this dependence is feeble.

In the framework of the QCD-improved parton model, quarks and gluons appear as
external particles, so that the knowledge of the decoupling constants
$\zeta_2$ and $\zeta_3$, which emerge as by-products of our analysis, is 
actually of practical interest for higher-order calculations.
In contrast to $\zeta_m$ and $\zeta_g$, $\zeta_2$ and $\zeta_3$ are $\xi$
dependent.
For future applications, we list $\zeta_2$ and $\zeta_3$ in
Appendix~\ref{appz2z3}.

\boldmath
\section{\label{seclet}Low-energy theorems for the $ggH$ and $q\bar qH$
interactions}
\unboldmath

The interactions of the SM Higgs boson with gluons and light quarks are
greatly affected by the virtual presence of the top quark.
In fact, the Higgs-boson coupling to gluons is essentially generated by a
top-quark loop alone.
In general, the theoretical description of such interactions is very 
complicated because there are two different mass scales involved, $M_H$ and 
$M_t$.
However, in the limit $M_H\ll2M_t$, the situation may be greatly simplified by
integrating out the top quark, i.e.\ by constructing a heavy-top-quark
effective Lagrangian.

The starting point of our consideration is the bare Yukawa Lagrangian of the
full theory,
\begin{equation}
{\cal L}=-\frac{H^0}{v^0}
\sum_{i=1}^{n_f}m_{q_i}^0\bar\psi_{q_i}^0\psi_{q_i}^0,
\end{equation}
which governs the interactions of the neutral CP-even Higgs boson $H$ with all
$n_f$ quark flavours, including the heavy one.
Here, $v$ is the Higgs vacuum expectation value.
The heavy-quark effective Lagrangian describing the interactions of $H$ with
the gluon $G$ and the $n_l$ light-quark flavours may be written in bare form
as
\begin{equation}
{\cal L}_{\rm eff}=-\frac{H^0}{v^0}\sum_{i=1}^5C_i^0{\cal O}_i^\prime.
\label{eqeff}
\end{equation}
The operators ${\cal O}_i^\prime$ are only constructed from light degrees of
freedom and read \cite{InaKubOka83,Klu75,Spi84}
\begin{eqnarray}
{\cal O}^\prime_1&\n=\n&\left(G^{0\prime,a}_{\mu\nu}\right)^2,
\nonumber\\
{\cal O}^\prime_2&\n=\n&
\sum_{i=1}^{n_l}m_{q_i}^{0\prime}\bar\psi_{q_i}^{0\prime}\psi_{q_i}^{0\prime},
\nonumber\\
{\cal O}^\prime_3&\n=\n&
\sum_{i=1}^{n_l}\bar\psi_{q_i}^{0\prime}
\left(i\not\!\!D^{0\prime}-m_{q_i}^{0\prime}\right)
\psi_{q_i}^{0\prime},
\nonumber\\
{\cal O}^\prime_4&\n=\n&G_\nu^{0\prime,a}
\left(\nabla^{ab}_\mu G^{0\prime,b\mu\nu}
+g_s^{0\prime}\sum_{i=1}^{n_l}\bar\psi_{q_i}^{0\prime}
\frac{\lambda^a}{2}\gamma^\nu\psi_{q_i}^{0\prime}\right)
-\partial_\mu \bar{c}^{0\prime,a}\partial^\mu c^{0\prime,a},
\nonumber\\
{\cal O}^\prime_5&\n=\n&
(\nabla^{ab}_\mu \partial^\mu \bar{c}^{0\prime,b}) c^{0\prime,a},
\label{eqopera}
\end{eqnarray}
where $G_{\mu\nu}^a$ is the colour field strength,
$D_\mu=\partial_\mu-i\mu^{2\varepsilon}g_s(\lambda^a/2)G_\mu^a$ and 
$\nabla_\mu^{ab}=\delta^{ab}\partial_\mu-i\mu^{2\varepsilon}g_sf^{abc}G_\mu^c$
are the covariant derivatives acting on the quark and gluon/ghost fields,
respectively, and $\lambda^a$ are the Gell-Mann matrices.
The residual dependence on the mass $m_h$ of the heavy quark $h$ is contained
in the coefficient functions $C_i^0$.

In phenomenological applications, one is mainly interested in the renormalized
coefficient functions $C_1$ and $C_2$, since only these contribute to physical
observables.
The renormalization of $C_i^0$ and ${\cal O}_i^\prime$ $(i=1,2)$ has been
explained in Ref.~\cite{CheKniSte96}.
For the reader's convenience, we repeat here the key results.
Denoting the renormalized operators by square brackets, we have
\cite{Klu75,Spi84}
\begin{eqnarray}
\left[{\cal O}_1^\prime\right]&\n=\n&
\left[1+2\left(\frac{\alpha_s^\prime\partial}{\partial\alpha_s^\prime}
\ln Z_g^\prime\right)\right]{\cal O}_1^\prime
-4\left(\frac{\alpha_s^\prime\partial}{\partial\alpha_s^\prime}
\ln Z_m^\prime\right){\cal O}_2^\prime,
\nonumber\\
\left[{\cal O}_2^\prime\right]&\n=\n&{\cal O}_2^\prime.
\end{eqnarray}
Note that ${\cal O}_3^\prime$, ${\cal O}_4^\prime$, and ${\cal O}_5^\prime$ do
not mix with ${\cal O}_1^\prime$ and ${\cal O}_2^\prime$.
On the other hand, the coefficient functions are renormalized according to
\begin{eqnarray}
C_1&\n=\n&\frac{1}{1+2(\alpha_s^\prime\partial/\partial\alpha_s^\prime)
\ln Z_g^\prime}C_1^0,
\label{eqc1ren}\\
C_2&\n=\n&\frac{4(\alpha_s^\prime\partial/\partial\alpha_s^\prime)
\ln Z_m^\prime}
{1+2(\alpha_s^\prime\partial/\partial\alpha_s^\prime)\ln Z_g^\prime}C_1^0
+C_2^0.
\end{eqnarray}     
Consequently, the physical part of ${\cal L}_{\rm eff}$ takes the form
\begin{equation}
{\cal L}_{\rm eff}^{\rm phys}=-\frac{H^0}{v^0}
\left(C_1\left[{\cal O}_1^\prime\right]
+C_2\left[{\cal O}_2^\prime\right]\right).
\label{eqleff}
\end{equation}
$C_i$ and $\left[{\cal O}_i^\prime\right]$ $(i=1,2)$ are individually finite,
but, with the exception of $\left[{\cal O}_2^\prime\right]$, they are not
separately renormalization-group (RG) invariant.
In Ref.~\cite{CheKniSte96}, a RG-improved version of Eq.~(\ref{eqleff}) has 
been constructed by exploiting the RG-invariance of the trace of the 
energy-momentum tensor.
The ratio $H^0/v^0$ receives a finite renormalization factor, which is of
${\cal O}(G_FM_t^2)$.
Its two- and three-loop QCD corrections have been found in Refs.~\cite{Kni94}
and \cite{KniSte95}, respectively.

The derivation of formulae to compute the coefficient functions is very
similar to the case of the decoupling relations.
As an example, let us consider the derivation of a formula involving $C_1^0$
and $C_4^0$.
The starting point is the 1PI Green function of two gluons which contains a
zero-momentum insertion of the composite operator
${\cal O}_h=m_h^0\bar h^0h^0$.
In momentum space, it reads in bare form
\begin{eqnarray}
\delta^{ab}\Gamma_{GG{\cal O}_h}^{0,\mu\nu}(p)
&\n=\n&i^2\int dxdy\,e^{ip\cdot(x-y)}
\left\langle TG^{0,a\mu}(x)G^{0,b\nu}(y){\cal O}_h(0)
\right\rangle^{\rm1PI}
\nonumber\\
&\n=\n&\delta^{ab}\left[-g^{\mu\nu}p^2\Gamma_{GG{\cal O}_h}^0(p^2)
+\mbox{terms proportional to $p^\mu p^\nu$}\right],
\label{eqggo1}
\end{eqnarray}
where $p$ denotes the four-momentum flowing along the gluon line.
In the limit $m_h\to\infty$, ${\cal O}_h$ may be written as a linear
combination of the effective operators given in Eq.~(\ref{eqopera}), so that
\begin{eqnarray}
\Gamma_{GG{\cal O}_h}^{0,\mu\nu}(p)
&\n=\n&\frac{i^2}{8}\int dxdy\,e^{ip\cdot(x-y)}
\left\langle TG^{0,a\mu}(x)G^{0,a\nu}(y)
\left(C_1^0{\cal O}^\prime_1 + C_4^0{\cal O}^\prime_4\right)
\right\rangle^{\rm1PI}+\ldots
\nonumber\\
&\n=\n&
\frac{i^2}{8}\zeta_3^0
\int dxdy\,e^{ip\cdot(x-y)}
\left\langle TG^{0\prime,a\mu}(x)G^{0\prime,a\nu}(y) 
\left(C_1^0{\cal O}_1^\prime + C_4^0{\cal O}_4^\prime\right)
\right\rangle^{\rm1PI}+\ldots
\nonumber\\
&\n=\n&
-g^{\mu\nu}p^2\zeta_3^0(-4 C_1^0 + 2 C_4^0)
\left(1+\mbox{higher orders}\right)+\ldots,
\label{eqggoh}
\end{eqnarray}
where the ellipses indicate terms of ${\cal O}(1/m_h)$ and terms proportional
to $p^\mu p^\nu$.
In the second step, we have used Eq.~(\ref{eqdec}) together with the fact that
$\Gamma_{GG{\cal O}_h}^{0,\mu\nu}(p)$ represents an amputated Green function.
If we consider the coefficients of the transversal part in the limit $p\to0$,
we observe that the contributions due to the higher-order corrections on the
r.h.s.\ of Eq.~(\ref{eqggoh}) vanish, as massless tadpoles are set to zero
in dimensional regularization.
The contribution due to the other operators vanish for the same reason.
On the l.h.s., only those diagrams survive which contain at least one
heavy-quark line.
Consequently, the hard part of the amputated Green function is given by
\begin{equation}
\Gamma_{GG{\cal O}_h}^{0h}(0)=\zeta_3^0\left(-4C_1^0+2C_4^0\right).
\label{eqamput}
\end{equation}
It is convenient to generate the diagrams contributing to the l.h.s.\ of 
Eq.~(\ref{eqamput}) by differentiating the gluon propagator w.r.t.\ the 
heavy-quark mass, so that we finally arrive at
\begin{equation}
\zeta_3^0(-4C_1^0+2C_4^0)= -\frac{1}{2}\partial_h^0 \Pi_G^{0h}(0),
\label{eqcfs0}
\end{equation}
where $\partial_h^0=\left[(m_h^0)^2\partial/\partial(m_h^0)^2\right]$.
The last equation results from the fact that the Yukawa coupling of the Higgs
boson to the heavy quark is proportional to $m_h^0$.

In a similar way, we obtain four more relationships, namely
\begin{eqnarray}
\zeta_2^0C_3^0&\n=\n&
-\frac{1}{2}\partial_h^0\Sigma_V^{0h}(0),
\nonumber\\
\zeta_m^0\zeta_2^0(C_2^0-C_3^0)&\n=\n&
1-\Sigma_S^{0h}(0)-\frac{1}{2}\partial_h^0\Sigma_S^{0h}(0),
\nonumber\\
\tilde\zeta_3^0(C_4^0+C_5^0)&\n=\n&\frac{1}{2}\partial_h^0\Pi_c^{0h}(0), 
\nonumber\\
\tilde\zeta_1^0C_5^0&\n=\n&\frac{1}{2}\partial_h^0\Gamma_{G\bar cc}^{0h}(0,0).
\label{eqcfs}
\end{eqnarray}
We may now solve Eqs.~(\ref{eqcfs0}) and (\ref{eqcfs}) for the coefficient
functions $C_i^0$ $(i=1,\ldots,5)$.

It is tempting to insert Eqs.~(\ref{eqdeccon0})--(\ref{eqdeccon2}) into
Eqs.~(\ref{eqcfs0}) and (\ref{eqcfs}), so as to express the bare coefficient
functions in terms of derivatives of the decoupling constants w.r.t.\ $m_h^0$.
Solving the three equations involving $C_1^0$, $C_4^0$, and $C_5^0$ for
$C_1^0$, we obtain
\begin{equation}
-2C_1^0=\partial_h^0\ln(\zeta_g^0)^2.
\end{equation}
Next, we express $\zeta_g^0$ through renormalized quantities.
Using $\partial_h^0=\partial_h$, we find
\begin{eqnarray}
-2C_1^0&\n=\n&\partial_h\ln(\zeta_g^0)^2
\nonumber\\
&\n=\n&\partial_h\ln\frac{\alpha_s^{0\prime}}{\alpha_s^0}
\nonumber\\
&\n=\n&\partial_h\ln(Z_g^\prime)^2+\partial_h\ln\alpha_s^\prime 
\nonumber\\
&\n=\n&\left[1+\frac{\alpha_s^\prime\partial}{\partial\alpha_s^\prime}
\ln\left(Z_g^\prime\right)^2\right]\partial_h\ln\alpha_s^\prime.
\end{eqnarray}
Identifying the renormalization factor of Eq.~(\ref{eqc1ren}), we obtain the
amazingly simple relation
\begin{eqnarray}
-2C_1&\n=\n&\partial_h\ln\alpha_s^\prime
\nonumber\\
&\n=\n&\partial_h\ln\zeta_g^2.
\label{eqc1let1}
\end{eqnarray}
This relation opens the possibility to compute $C_1$ through
${\cal O}(\alpha_s^4)$, since one only needs to know the logarithmic pieces of
$\zeta_g$ in this order, which may be reconstructed from its lower-order terms
in combination with the four-loop $\beta$ \cite{LarRitVer971} and $\gamma_m$
\cite{Che97LarRitVer972} functions.
It is possible to directly relate $C_1$ to the $\beta$ and $\gamma_m$
functions of the full and effective theories.
Exploiting the relation
\begin{eqnarray}
\beta^\prime(\alpha_s^\prime) 
&\n=\n&\frac{\mu^2d}{d\mu^2}\,\frac{\alpha_s^\prime}{\pi}
\nonumber\\
&\n=\n&\left[\frac{\mu^2\partial}{\partial\mu^2}
+\beta(\alpha_s)\frac{\partial}{\partial\alpha_s}
+\gamma_m(\alpha_s)\frac{m_h\partial}{\partial m_h}
\right]\frac{\alpha_s^\prime}{\pi},
\end{eqnarray}
where $\alpha_s^\prime=\alpha_s^\prime(\mu,\alpha_s,m_h)$, we find
\begin{equation}
C_1=\frac{\pi}{2\alpha_s^\prime\left[1-2\gamma_m(\alpha_s)\right]}
\left[\beta^\prime(\alpha_s^\prime)
-\beta(\alpha_s)\frac{\partial\alpha_s^\prime}{\partial\alpha_s}\right].
\label{eqc1let}
\end{equation}
In the case of $C_2$, we may proceed along the same lines to obtain
\begin{eqnarray}
C_2&\n=\n&1+2\partial_h\ln\zeta_m
\nonumber\\
&\n=\n&1-\frac{2}{1-2\gamma_m(\alpha_s)}
\left[\gamma_m^\prime(\alpha_s^\prime)-\gamma_m(\alpha_s)
-\beta(\alpha_s)\frac{1}{m_q^\prime}\,
\frac{\partial m_q^\prime}{\partial\alpha_s}\right],
\label{eqc2let}
\end{eqnarray}
where $m_q^\prime=m_q^\prime(\mu,\alpha_s,m_h)$.
It should be stressed that Eqs.~(\ref{eqc1let}) and (\ref{eqc2let}) are valid
to all orders in $\alpha_s$.

Fully exploiting present knowledge of the $\beta$ \cite{LarRitVer971} and
$\gamma_m$ \cite{Che97LarRitVer972} functions, we may evaluate $C_1$ and $C_2$
through ${\cal O}(\alpha_s^4)$ via Eqs.~(\ref{eqc1let}) and (\ref{eqc2let}).
In the pure $\overline{\mbox{MS}}$ scheme, we find
\begin{eqnarray}
C_1&\n=\n&
-\frac{1}{12}\,\frac{\alpha_s^{(n_f)}(\mu)}{\pi}
\Bigg\{
  1 
+ \frac{\alpha_s^{(n_f)}(\mu)}{\pi}
\Bigg(
\frac{11}{4} 
- \frac{1}{6} \ln\frac{\mu^2}{m_h^2}
\Bigg)
\nonumber\\
&\n\n&{}+ \left(\frac{\alpha_s^{(n_f)}(\mu)}{\pi}\right)^2
\Bigg[
\frac{2821}{288} 
- \frac{3}{16} \ln\frac{\mu^2}{m_h^2}
+ \frac{1}{36} \ln^2\frac{\mu^2}{m_h^2}
+ n_l\left(
-\frac{67}{96} 
+ \frac{1}{3} \ln\frac{\mu^2}{m_h^2}
\right)
\Bigg]
\nonumber\\
&\n\n&{}+ \left(\frac{\alpha_s^{(n_f)}(\mu)}{\pi}\right)^3
\Bigg[
-\frac{4004351}{62208} 
+ \frac{1305893}{13824}\zeta(3)
- \frac{859}{288} \ln\frac{\mu^2}{m_h^2}
+ \frac{431}{144} \ln^2\frac{\mu^2}{m_h^2}
- \frac{1}{216} \ln^3\frac{\mu^2}{m_h^2}
\nonumber\\
&\n\n&{}+  n_l \left(
  \frac{115607}{62208} 
- \frac{110779}{13824}\zeta(3)
+ \frac{641}{432} \ln\frac{\mu^2}{m_h^2}
+ \frac{151}{288} \ln^2\frac{\mu^2}{m_h^2}
\right) 
\nonumber\\
&\n\n&{}+ n_l^2 \left(
- \frac{6865}{31104} 
+ \frac{77}{1728} \ln\frac{\mu^2}{m_h^2} 
- \frac{1}{18} \ln^2\frac{\mu^2}{m_h^2}
\right)
\Bigg]
\Bigg\}
\nonumber\\
&\n\approx\n&
-\frac{1}{12}\,\frac{\alpha_s^{(n_f)}(\mu_h)}{\pi}
\Bigg[1
+ 2.7500
\frac{\alpha_s^{(n_f)}(\mu_h)}{\pi}
+ \left(9.7951 - 0.6979\,n_l\right) 
\left(\frac{\alpha_s^{(n_f)}(\mu_h)}{\pi}\right)^2
\nonumber\\
&\n\n&
+ \left(49.1827 - 7.7743\,n_l - 0.2207\,n_l^2 \right)
\left(\frac{\alpha_s^{(n_f)}(\mu_h)}{\pi}\right)^3\Bigg],
\label{eqc1}\\
C_2 &\n=\n&
1 
+ \left(\frac{\alpha_s^{(n_f)}(\mu)}{\pi}\right)^2 \Bigg(
  \frac{5}{18} 
- \frac{1}{3} \ln\frac{\mu^2}{m_h^2}
\Bigg)
\nonumber\\
&\n\n&{}+ \left(\frac{\alpha_s^{(n_f)}(\mu)}{\pi}\right)^3 \Bigg[
  \frac{311}{1296} 
+ \frac{5}{3}\zeta(3)
- \frac{175}{108} \ln\frac{\mu^2}{m_h^2}
- \frac{29}{36} \ln^2\frac{\mu^2}{m_h^2}
+ n_l\left(
  \frac{53}{216} 
+ \frac{1}{18} \ln^2\frac{\mu^2}{m_h^2}
\right)
\Bigg]
\nonumber\\
&\n\n&{}+ \left(\frac{\alpha_s^{(n_f)}(\mu)}{\pi}\right)^4 \Bigg[
  \frac{2800175}{186624} 
+ \frac{373261}{13824}\zeta(3) 
- \frac{155}{6}\zeta(4)
- \frac{575}{36}\zeta(5)
+ \frac{31}{72}B_4
\nonumber\\
&\n\n&{}+ \left( 
  -\frac{50885}{2592} 
+ \frac{155}{12}\zeta(3) 
\right) \ln\frac{\mu^2}{m_h^2}
- \frac{1219}{216} \ln^2\frac{\mu^2}{m_h^2}
- \frac{301}{144} \ln^3\frac{\mu^2}{m_h^2}
\nonumber\\
&\n\n&{}+ n_l \left(
- \frac{16669}{15552} 
- \frac{221}{288}\zeta(3)
+ \frac{25}{12}\zeta(4)
- \frac{1}{36} B_4
+ \frac{7825}{2592} \ln\frac{\mu^2}{m_h^2} 
+ \frac{23}{48} \ln^2\frac{\mu^2}{m_h^2}
+ \frac{5}{18} \ln^3\frac{\mu^2}{m_h^2}
\right) 
\nonumber\\
&\n\n&{}+  n_l^2 \left(
  \frac{3401}{23328} 
- \frac{7}{54}\zeta(3)
- \frac{31}{324} \ln\frac{\mu^2}{m_h^2}
- \frac{1}{108} \ln^3\frac{\mu^2}{m_h^2}
\right)
\Bigg]
\nonumber\\
&\n\approx\n&
1
+ 0.2778
\left(\frac{\alpha_s^{(n_f)}(\mu_h)}{\pi}\right)^2
+ \left(2.2434 + 0.2454\,n_l\right) 
\left(\frac{\alpha_s^{(n_f)}(\mu_h)}{\pi}\right)^3
\nonumber\\
&\n\n&
+ \left(2.1800 + 0.3096\,n_l - 0.0100\,n_l^2 \right)
\left(\frac{\alpha_s^{(n_f)}(\mu_h)}{\pi}\right)^4,
\label{eqc2}
\end{eqnarray}
where, for simplicity, we have chosen $\mu=\mu_h$ in the approximate
expressions.
The ${\cal O}(\alpha_s^3)$ term of Eq.~(\ref{eqc2}) may also be found in
Ref.~\cite{CheSte97}.
The corresponding expressions written with the pole mass $M_h$ read
\begin{eqnarray}
C_1^{\rm OS}&\n=\n&
-\frac{1}{12}\,\frac{\alpha_s^{(n_f)}(\mu)}{\pi}
\Bigg\{1 
+ \frac{\alpha_s^{(n_f)}(\mu)}{\pi}
\Bigg(
\frac{11}{4} 
- \frac{1}{6} \ln\frac{\mu^2}{M_h^2}
\Bigg)
\nonumber\\
&\n\n&{}+ \left(\frac{\alpha_s^{(n_f)}(\mu)}{\pi}\right)^2
\Bigg[
\frac{2693}{288} 
- \frac{25}{48} \ln\frac{\mu^2}{M_h^2}
+ \frac{1}{36} \ln^2\frac{\mu^2}{M_h^2}
+ n_l\left(
-\frac{67}{96} 
+ \frac{1}{3} \ln\frac{\mu^2}{M_h^2}
\right)
\Bigg]
\nonumber\\
&\n\n&{}+ \left(\frac{\alpha_s^{(n_f)}(\mu)}{\pi}\right)^3
\Bigg[
-\frac{4271255}{62208} 
-\frac{2}{3}\zeta(2)\left(1+\frac{\ln2}{3}\right)
+ \frac{1306661}{13824}\zeta(3)
\nonumber\\
&\n\n&{}- \frac{4937}{864} \ln\frac{\mu^2}{M_h^2}
+ \frac{385}{144} \ln^2\frac{\mu^2}{M_h^2}
- \frac{1}{216} \ln^3\frac{\mu^2}{M_h^2}
\nonumber\\
&\n\n&{}+  n_l \left(
  \frac{181127}{62208}
+ \frac{1}{9}\zeta(2) 
- \frac{110779}{13824}\zeta(3)
+ \frac{109}{48} \ln\frac{\mu^2}{M_h^2}
+ \frac{53}{96} \ln^2\frac{\mu^2}{M_h^2}
\right) 
\nonumber\\
&\n\n&{}+ n_l^2 \left(
- \frac{6865}{31104} 
+ \frac{77}{1728} \ln\frac{\mu^2}{M_h^2} 
- \frac{1}{18} \ln^2\frac{\mu^2}{M_h^2}
\right)
\Bigg]
\Bigg\}
\nonumber\\
&\n\approx\n&
-\frac{1}{12}\,\frac{\alpha_s^{(n_f)}(M_h)}{\pi}
\Bigg[1
+ 2.7500
\frac{\alpha_s^{(n_f)}(M_h)}{\pi}
+ \left(9.3507 - 0.6979\,n_l\right) 
\left(\frac{\alpha_s^{(n_f)}(M_h)}{\pi}\right)^2
\nonumber\\
&\n\n&
+ \left(43.6090 - 6.5383\,n_l - 0.2207\,n_l^2 \right)
\left(\frac{\alpha_s^{(n_f)}(M_h)}{\pi}\right)^3\Bigg],
\label{eqc1os}\\
C_2^{\rm OS}&\n=\n&1 
+ \left(\frac{\alpha_s^{(n_f)}(\mu)}{\pi}\right)^2 \Bigg(
  \frac{5}{18} 
- \frac{1}{3} \ln\frac{\mu^2}{M_h^2}
\Bigg)
\nonumber\\
&\n\n&{}
+\left(\frac{\alpha_s^{(n_f)}(\mu)}{\pi}\right)^3 \Bigg[
- \frac{841}{1296} 
+ \frac{5}{3}\zeta(3)
- \frac{247}{108} \ln\frac{\mu^2}{M_h^2}
- \frac{29}{36} \ln^2\frac{\mu^2}{M_h^2}
\nonumber\\
&\n\n&{}
+ n_l\left(
  \frac{53}{216} 
+ \frac{1}{18} \ln^2\frac{\mu^2}{M_h^2}
\right)
\Bigg]
\nonumber\\
&\n\n&{}+ \left(\frac{\alpha_s^{(n_f)}(\mu)}{\pi}\right)^4 \Bigg[
  \frac{578975}{186624} 
- \frac{4}{3}\zeta(2)\left(1+\frac{\ln2}{3}\right)
+ \frac{374797}{13824}\zeta(3) 
- \frac{155}{6}\zeta(4)
- \frac{575}{36}\zeta(5)
\nonumber\\
&\n\n&{}
+ \frac{31}{72} B_4
+ \left(
  -\frac{83405}{2592}
+ \frac{155}{12} \zeta(3) 
\right) \ln\frac{\mu^2}{M_h^2}
- \frac{2101}{216} \ln^2\frac{\mu^2}{M_h^2}
- \frac{301}{144} \ln^3\frac{\mu^2}{M_h^2}
\nonumber\\
&\n\n&{}+ n_l \left(
- \frac{11557}{15552} 
+ \frac{2}{9}\zeta(2)
- \frac{221}{288}\zeta(3)
+ \frac{25}{12}\zeta(4)
- \frac{1}{36} B_4
\right.\nonumber\\
&\n\n&{}+\left.
 \frac{9217}{2592} \ln\frac{\mu^2}{M_h^2}
+ \frac{109}{144} \ln^2\frac{\mu^2}{M_h^2}
+ \frac{5}{18} \ln^3\frac{\mu^2}{M_h^2}
\right)
\nonumber\\
&\n\n&{}+ n_l^2 \left(
  \frac{3401}{23328} 
- \frac{7}{54}\zeta(3)
- \frac{31}{324} \ln\frac{\mu^2}{M_h^2}
- \frac{1}{108} \ln^3\frac{\mu^2}{M_h^2}
\right)
\Bigg]
\nonumber\\
&\n\approx\n&1
+ 0.2778
\left(\frac{\alpha_s^{(n_f)}(M_h)}{\pi}\right)^2
+ \left(1.3545 + 0.2454\,n_l \right)
\left(\frac{\alpha_s^{(n_f)}(M_h)}{\pi}\right)^3
\nonumber\\
&\n\n&
+ \left(-12.2884 + 1.0038\,n_l - 0.0100\,n_l^2 \right)
\left(\frac{\alpha_s^{(n_f)}(M_h)}{\pi}\right)^4,
\label{eqc2os}
\end{eqnarray}
where we have put $\mu=M_h$ in the numerical evaluations.
The ${\cal O}(\alpha_s^3)$ term of Eq.~(\ref{eqc1os}) may also be found in
Ref.~\cite{CheKniSte97}, where the hadronic decay width of the SM Higgs boson
has been calculated through ${\cal O}(\alpha_s^4)$.

\boldmath
\section{\label{secgam}Low-energy theorem for the $\gamma\gamma H$
interaction}
\unboldmath

In this section, we extend the formalism developed in Sections~\ref{seccon}
and \ref{seclet} to include the Higgs-boson interactions with photons.
From the technical point of view, this means that we are now concerned with
QCD corrections to Green functions where the external particles are photons
instead of gluons.
In the following, this is indicated by an additional subscript `$\gamma$'.
Furthermore, we do not need to consider Green functions involving external
ghost lines any more.
The Abelian versions of Eqs.~(\ref{eqren}), (\ref{eqdec}), and (\ref{eqopera})
emerge via the substitutions
$g\to\bar e=\sqrt{4\pi\bar\alpha}$,
$G_\mu^a\to A_\mu$,
$G_{\mu\nu}^a\to F_{\mu\nu}$,
$\lambda^a/2\to 1$, and
$f^{abc}\to0$,
where $\bar e$ is the gauge coupling of quantum electrodynamics (QED),
$\bar\alpha$ is the fine-structure constant,
$A_\mu$ is the photon field, and
$F_{\mu\nu}$ is the electromagnetic field strength.
We continue to work in the $\overline{\mbox{MS}}$ renormalization scheme.
The system of composite operators~(\ref{eqopera}) is reduced to the Abelian
counterparts of the gauge-invariant operators ${\cal O}_1^\prime$,
${\cal O}_2^\prime$, and ${\cal O}_3^\prime$.
The other formulae derived in Sections~\ref{seccon} and \ref{seclet} simplify 
accordingly.

Similarly to Eq.~(\ref{eqdecg}), the decoupling relation for the renormalized
$\overline{\mbox{MS}}$ fine-structure constant reads
\begin{equation}
\bar\alpha^\prime(\mu)=
\left(\frac{Z_{g\gamma}}{Z^\prime_{g\gamma}}\zeta_{g\gamma}^0\right)^2
\bar\alpha(\mu)
=\zeta_{g\gamma}^2 \bar\alpha(\mu).
\end{equation}
Using
\begin{equation}
Z_{g\gamma}=\frac{1}{\sqrt{Z_{3\gamma}}},\qquad
\zeta_{g\gamma}^0=\frac{1}{\sqrt{\zeta_{3\gamma}^0}},
\end{equation}
we thus obtain
\begin{equation}
\zeta_{g\gamma}^2=\frac{Z_{3\gamma}^\prime}{Z_{3\gamma}\zeta_{3\gamma}^0}.
\end{equation}
To three loops in QCD, the photon wave-function-renormalization constant
$Z_{3\gamma}$ may be extracted from Ref.~\cite{CheKatTka79}, while the bare
decoupling constant $\zeta_{g\gamma}^0$ for $\bar\alpha(\mu)$ is determined by
the hard part of the three-loop photon propagator, which may be found in
Refs.~\cite{CheKueSte96,Mat96}.
Putting everything together, we find in the pure $\overline{\mbox{MS}}$ scheme
\begin{eqnarray}
\zeta_{g\gamma}^2&\n=\n&1
+\frac{\bar\alpha^{(n_f)}(\mu)}{\pi}\Bigg\{
-Q_h^2\ln\frac{\mu^2}{m_h^2}
+\frac{\alpha_s^{(n_f)}(\mu)}{\pi}Q_h^2\left(
-\frac{13}{12}
+\ln\frac{\mu^2}{m_h^2}\right)
\nonumber\\
&\n\n&{}
+\left(\frac{\alpha_s^{(n_f)}(\mu)}{\pi}\right)^2\Bigg[
Q_h^2\Bigg(
\frac{10819}{2592} 
-\frac{655}{144}\zeta(3)
+\frac{8}{9}\ln\frac{\mu^2}{m_h^2}
+\frac{31}{24}\ln^2\frac{\mu^2}{m_h^2}
\nonumber\\
&\n\n&{}
+n_l\left(
-\frac{361}{1296}
+\frac{1}{18}\ln\frac{\mu^2}{m_h^2}
-\frac{1}{12}\ln^2\frac{\mu^2}{m_h^2}\right)
\Bigg)
\nonumber\\
&\n\n&{}
+\sum_{i=1}^{n_l}Q_{q_i}^2\left(
-\frac{295}{1296} 
+\frac{11}{72}\ln\frac{\mu^2}{m_h^2}
-\frac{1}{12}\ln^2\frac{\mu^2}{m_h^2}\right)
\Bigg]\Bigg\},
\label{eqzetagg}
\end{eqnarray}
where $Q_q$ is the fractional electric charge of quark flavour $q$.
Introducing the pole mass $M_h$, this becomes
\begin{eqnarray}
\left(\zeta_{g\gamma}^{\rm OS}\right)^2&\n=\n&1
+\frac{\bar\alpha^{(n_f)}(\mu)}{\pi}\Bigg\{
-Q_h^2\ln\frac{\mu^2}{M_h^2}
+\frac{\alpha_s^{(n_f)}(\mu)}{\pi}Q_h^2\left(
-\frac{15}{4} 
-\ln\frac{\mu^2}{M_h^2}\right)
\nonumber\\
&\n\n&{}
+\left(\frac{\alpha_s^{(n_f)}(\mu)}{\pi}\right)^2\Bigg[
Q_h^2\Bigg(
-\frac{41219}{2592} 
-4\zeta(2)\left(1+\frac{\ln2}{3}\right)
-\frac{607}{144}\zeta(3)
\nonumber\\
&\n\n&{}
-\frac{437}{36}\ln\frac{\mu^2}{M_h^2}
-\frac{31}{24}\ln^2\frac{\mu^2}{M_h^2}
+n_l\left(
\frac{917}{1296}
+\frac{2}{3}\zeta(2)
+\frac{7}{9}\ln\frac{\mu^2}{M_h^2}
+\frac{1}{12}\ln^2\frac{\mu^2}{M_h^2}\right)
\Bigg)
\nonumber\\
&\n\n&{}
+\sum_{i=1}^{n_l}Q_{q_i}^2\left(
-\frac{295}{1296}
+\frac{11}{72}\ln\frac{\mu^2}{M_h^2}
-\frac{1}{12}\ln^2\frac{\mu^2}{M_h^2}\right)
\Bigg]\Bigg\}.
\label{eqzetaggos}
\end{eqnarray}

Next, we turn to the coefficient function $C_{1\gamma}$ of the heavy-quark
effective $\gamma\gamma H$ coupling.
Equation~(\ref{eqc1let}) undergoes obvious modifications to become
\begin{eqnarray}
C_{1\gamma}&\n=\n&
-\frac{1}{2}\partial_h\ln\zeta_{g\gamma}^2
\nonumber\\
&\n=\n&
\frac{\pi}{2\bar\alpha^\prime\left[1-2\gamma_m(\alpha_s)\right]}
\left[\beta^\prime_\gamma(\bar\alpha^\prime,\alpha_s^\prime)
-\beta_\gamma(\bar\alpha,\alpha_s)
\frac{\partial\bar\alpha^\prime}{\partial\bar\alpha}
-\beta(\alpha_s)
\frac{\partial\bar\alpha^\prime}{\partial\alpha_s}\right],
\end{eqnarray}
where $\beta_\gamma$ is the $\beta$ function governing the running of the
$\overline{\mbox{MS}}$ fine-structure constant $\bar\alpha(\mu)$.
An expression for $\beta_\gamma$ may be extracted from the well-known QCD
corrections to the photon propagator \cite{GorKatLar91}.
Restricting ourselves to the leading order in $\bar\alpha$, we find
\begin{equation}
\frac{\mu^2d}{d\mu^2}\,\frac{\bar\alpha^{(n_f)}}{\pi}
=\beta_\gamma^{(n_f)}\left(\bar\alpha^{(n_f)},\alpha_s^{(n_f)}\right)
=\left(\frac{\bar\alpha^{(n_f)}}{\pi}\right)^2
\sum_{N=1}^\infty\beta_{N-1,\gamma}^{(n_f)}
\left(\frac{\alpha_s^{(n_f)}}{\pi}\right)^{N-1},
\end{equation}
where $N$ denotes the number of loops and
\begin{eqnarray}
\beta_{0\gamma}^{(n_f)}&\n=\n&\beta_{1\gamma}^{(n_f)}
=\sum_{i=1}^{n_f}Q_{q_i}^2,
\nonumber\\
\beta_{2\gamma}^{(n_f)}&\n=\n&
\frac{1}{64}\sum_{i=1}^{n_f}Q_{q_i}^2
\left(\frac{500}{3}-\frac{88}{9}n_f\right),
\nonumber\\
\beta_{3\gamma}^{(n_f)}&\n=\n&\frac{1}{256}\Bigg\{
\sum_{i=1}^{n_f}Q_{q_i}^2\left[
\frac{41948}{27}+\frac{7040}{9}\zeta(3)
+n_f\left(-\frac{5656}{27}-\frac{7040}{27}\zeta(3)\right)
-\frac{1232}{243}n_f^2\right]
\nonumber\\
&\n\n&{}
+\left(\sum_{i=1}^{n_f}Q_{q_i}\right)^2\left(
\frac{1760}{27}-\frac{1280}{9}\zeta(3)\right)
\Bigg\}.
\end{eqnarray}
We are now in the position to evaluate $C_{1\gamma}$ up to the four-loop
level.
In the pure $\overline{\mbox{MS}}$ scheme, we find
\begin{eqnarray}
C_{1\gamma}&\n=\n& 
-\frac{1}{2}\,\frac{\bar\alpha^{(n_f)}(\mu)}{\pi}\Bigg\{
Q_h^2
-\frac{\alpha_s^{(n_f)}(\mu)}{\pi}Q_h^2
+\left(\frac{\alpha_s^{(n_f)}(\mu)}{\pi}\right)^2\Bigg[
Q_h^2\Bigg(
-\frac{8}{9} 
-\frac{31}{12}\ln\frac{\mu^2}{m_h^2}
\nonumber\\
&\n\n&{}
+n_l\left(
-\frac{1}{18}
+\frac{1}{6}\ln\frac{\mu^2}{m_h^2}\right)\Bigg)
+\sum_{i=1}^{n_l}Q_{q_i}^2\left( 
-\frac{11}{72}
+\frac{1}{6}\ln\frac{\mu^2}{m_h^2}\right)\Bigg]
\nonumber\\
&\n\n&{}
+\left(\frac{\alpha_s^{(n_f)}(\mu)}{\pi}\right)^3\Bigg[
Q_h^2\Bigg(
-\frac{95339}{2592}
+\frac{7835}{288}\zeta(3)
-\frac{541}{108}\ln\frac{\mu^2}{m_h^2}
-\frac{961}{144}\ln^2\frac{\mu^2}{m_h^2}
\nonumber\\
&\n\n&{}
+n_l\left(
\frac{4693}{1296} 
-\frac{125}{144}\zeta(3)
+\frac{101}{216}\ln\frac{\mu^2}{m_h^2}
+\frac{31}{36}\ln^2\frac{\mu^2}{m_h^2}\right)
\nonumber\\
&\n\n&{}
+n_l^2\left(
-\frac{19}{324} 
+\frac{1}{54}\ln\frac{\mu^2}{m_h^2}
-\frac{1}{36}\ln^2\frac{\mu^2}{m_h^2}\right)\Bigg)
\nonumber\\
&\n\n&{}
+\sum_{i=1}^{n_l}Q_{q_i}^2\left(
\frac{53}{108}
-\frac{55}{54}\zeta(3)
+\frac{11}{54}\ln\frac{\mu^2}{m_h^2}
+\frac{29}{72}\ln^2\frac{\mu^2}{m_h^2}
+ n_l\left(
-\frac{449}{3888} 
-\frac{1}{36}\ln^2\frac{\mu^2}{m_h^2}\right)\right)
\nonumber\\
&\n\n&{}
+\left(\left(\sum_{i=1}^{n_f}Q_{q_i}\right)^2
-\left(\sum_{i=1}^{n_l}Q_{q_i}\right)^2\right)
\left(\frac{55}{216}-\frac{5}{9}\zeta(3)\right)\Bigg]\Bigg\}.
\label{eqc1gamma}
\end{eqnarray}
The corresponding expression written with the pole mass $M_h$ reads
\begin{eqnarray}
C_{1\gamma}^{\rm OS}&\n=\n&
-\frac{1}{2}\,\frac{\bar\alpha^{(n_f)}(\mu)}{\pi}\Bigg\{
Q_h^2
-\frac{\alpha_s^{(n_f)}(\mu)}{\pi}Q_h^2 
+\left(\frac{\alpha_s^{(n_f)}(\mu)}{\pi}\right)^2\Bigg[
Q_h^2\Bigg(
-\frac{8}{9} 
-\frac{31}{12}\ln\frac{\mu^2}{M_h^2}
\nonumber\\
&\n\n&{}
+n_l\left(
-\frac{1}{18}
+ \frac{1}{6} \ln\frac{\mu^2}{M_h^2}\right)\Bigg)
+\sum_{i=1}^{n_l}Q_{q_i}^2\left(
-\frac{11}{72}
+\frac{1}{6}\ln\frac{\mu^2}{M_h^2}\right)\Bigg]
\nonumber\\
&\n\n&{}
+\left(\frac{\alpha_s^{(n_f)}(\mu)}{\pi}\right)^3\Bigg[
Q_h^2\Bigg(
-\frac{113195}{2592} 
+\frac{7835}{288}\zeta(3)
-\frac{1099}{108}\ln\frac{\mu^2}{M_h^2}
-\frac{961}{144}\ln^2\frac{\mu^2}{M_h^2}
\nonumber\\
&\n\n&{}
+n_l\left(
\frac{5269}{1296} 
-\frac{125}{144}\zeta(3)
+\frac{173}{216}\ln\frac{\mu^2}{M_h^2}
+\frac{31}{36}\ln^2\frac{\mu^2}{M_h^2}\right)
\nonumber\\
&\n\n&{}
+n_l^2\left(
-\frac{19}{324} 
+\frac{1}{54}\ln\frac{\mu^2}{M_h^2}
-\frac{1}{36}\ln^2\frac{\mu^2}{M_h^2}\right)\Bigg)
\nonumber\\
&\n\n&{}
+\sum_{i=1}^{n_l}Q_{q_i}^2\left(
\frac{101}{108} 
-\frac{55}{54}\zeta(3)
+\frac{29}{54}\ln\frac{\mu^2}{M_h^2}
+\frac{29}{72}\ln^2\frac{\mu^2}{M_h^2}
+n_l\left(
-\frac{449}{3888} 
-\frac{1}{36}\ln^2\frac{\mu^2}{M_h^2}\right)\right)
\nonumber\\
&\n\n&{}
+\left(\left(\sum_{i=1}^{n_f}Q_{q_i}\right)^2
-\left(\sum_{i=1}^{n_l}Q_{q_i}\right)^2\right)
\left(\frac{55}{216}-\frac{5}{9}\zeta(3)\right)\Bigg]\Bigg\}.
\label{eqc1gammaos}
\end{eqnarray}
The ${\cal O}(\bar\alpha\alpha_s)$ term of Eq.~(\ref{eqc1gammaos}) agrees with
Ref.~\cite{ZhengWu90}.
By comparing Eqs.~(\ref{eqc1gamma}) and (\ref{eqc1gammaos}), we notice that
only the ${\cal O}(\bar\alpha\alpha_s^3)$ terms depend on the definition of
the top quark mass.
The ${\cal O}(\bar\alpha\alpha_s^2)$ contributions due to the diagrams where
the Higgs boson and the photons are connected to the same quark loop agree
with Ref.~\cite{Mat96}, where the corresponding vertex diagrams have been
directly calculated.
In Ref.~\cite{Mat96}, also power-suppressed terms of
${\cal O}(\bar\alpha\alpha_s^2)$ have been considered.

\section{\label{secdis}Discussion and conclusions}

In the context of QCD with $\overline{\mbox{MS}}$ renormalization, the
decoupling constants $\zeta_m$ and $\zeta_g$ given in Eqs.~(\ref{eqzetam}) and
(\ref{eqzetag}), respectively, determine the shifts in the light-quark masses
$m_q(\mu)$ and the strong coupling constant $\alpha_s(\mu)$ that occur as the
threshold of a heavy-quark flavour $h$ is crossed.
In Eqs.~(\ref{eqzetam}) and (\ref{eqzetag}), the matching scale $\mu$, at
which the crossing is implemented, is measured in units of the
$\overline{\mbox{MS}}$ mass $m_h(\mu)$ of $h$.
The corresponding decoupling relations, formulated for the pole mass $M_h$, 
are given in Eqs.~(\ref{eqzetamos}) and (\ref{eqzetagos}).
Equations~(\ref{eqzetam}) and (\ref{eqzetag}) are valid through three loops,
i.e.\ they extend the results of Refs.~\cite{BerWet82,LarRitVer95} by one 
order.
These results will be indispensible in order to relate the QCD predictions for
different observables at next-to-next-to-next-to-leading order.
Meaningful estimates of such corrections already exist \cite{Sam95}.

We wish to stress that our calculation is based on a conceptually new approach
that directly links the decoupling constants to massive tadpole integrals.
This is crucial because the presently available analytic technology does not
permit the extension of the methods employed in
Refs.~\cite{BerWet82,LarRitVer95} to the order under consideration here.
In fact, this would require the evaluation of three-loop diagrams with
nonvanishing external momenta in the case of Ref.~\cite{BerWet82}, or even
four-loop diagrams in the case of Ref.~\cite{LarRitVer95}.

We are now in a position to explore the phenomenological implications of our 
results.
To that end, it is convenient to have perturbative solutions of
Eqs.~(\ref{eqrgea}) and (\ref{eqrgem}) for fixed $n_f$ in closed form.
Iteratively solving Eq.~(\ref{eqrgea}) yields \cite{CheKniSte97als}
\begin{eqnarray}
\frac{\alpha_s(\mu)}{\pi}&\n=\n&\frac{1}{\beta_0L}
-\frac{b_1\ln L}{(\beta_0L)^2}
+\frac{1}{(\beta_0L)^3}\left[b_1^2(\ln^2L-\ln L-1)+b_2\right]
\nonumber\\
&\n\n&+\frac{1}{(\beta_0L)^4}\left[
b_1^3\left(-\ln^3L+\frac{5}{2}\ln^2L+2\ln L-\frac{1}{2}\right)
-3b_1b_2\ln L+\frac{b_3}{2}\right],
\label{eqalp}
\end{eqnarray}
where $b_N=\beta_N/\beta_0$ ($N=1,2,3$), $L=\ln(\mu^2/\Lambda^2)$, and
terms of ${\cal O}(1/L^5)$ have been neglected.
The asymptotic scale parameter, $\Lambda$, is defined in the canonical way, by
demanding that Eq.~(\ref{eqalp}) does not contain a term proportional to
$({\rm const.}/L^2)$ \cite{BarBurDukMut78}.
Equation~(\ref{eqalp}) extends Eq.~(9.5a) of Ref.~\cite{Bar96} to four loops.
Combining Eqs.~(\ref{eqrgea}) and (\ref{eqrgem}), we obtain a differential 
equation for $m_q(\mu)$ as a function of $\alpha_s(\mu)$.
It has the solution \cite{GraBroGraSch90}
\begin{equation}
\frac{m_q(\mu)}{m_q(\mu_0)}=
\frac{c(\alpha_s(\mu)/\pi)}{c(\alpha_s(\mu_0)/\pi)},
\end{equation}
with \cite{Che97LarRitVer972}
\begin{eqnarray}
c(x)&\n=\n&x^{c_0}\left\{1+(c_1-b_1c_0)x
+\frac{1}{2}\left[(c_1-b_1c_0)^2+c_2-b_1c_1+b_1^2c_0-b_2c_0\right]x^2\right.
\nonumber\\
&\n\n&{}+\left[\frac{1}{6}(c_1-b_1c_0)^3
+\frac{1}{2}(c_1-b_1c_0)\left(c_2-b_1c_1+b_1^2c_0-b_2c_0\right)\right.
\nonumber\\
&\n\n&{}+\left.\left.
\frac{1}{3}\left(c_3-b_1c_2+b_1^2c_1-b_2c_1-b_1^3c_0+2b_1b_2c_0-b_3c_0\right)
\right]x^3\right\},
\label{eqmas}
\end{eqnarray}
where $c_N=\gamma_N/\beta_0$ ($N=0,\ldots,3$) and terms of ${\cal O}(x^4)$
have been neglected.

\begin{figure}[t]
\epsfxsize=16cm
\epsffile[90 275 463 579]{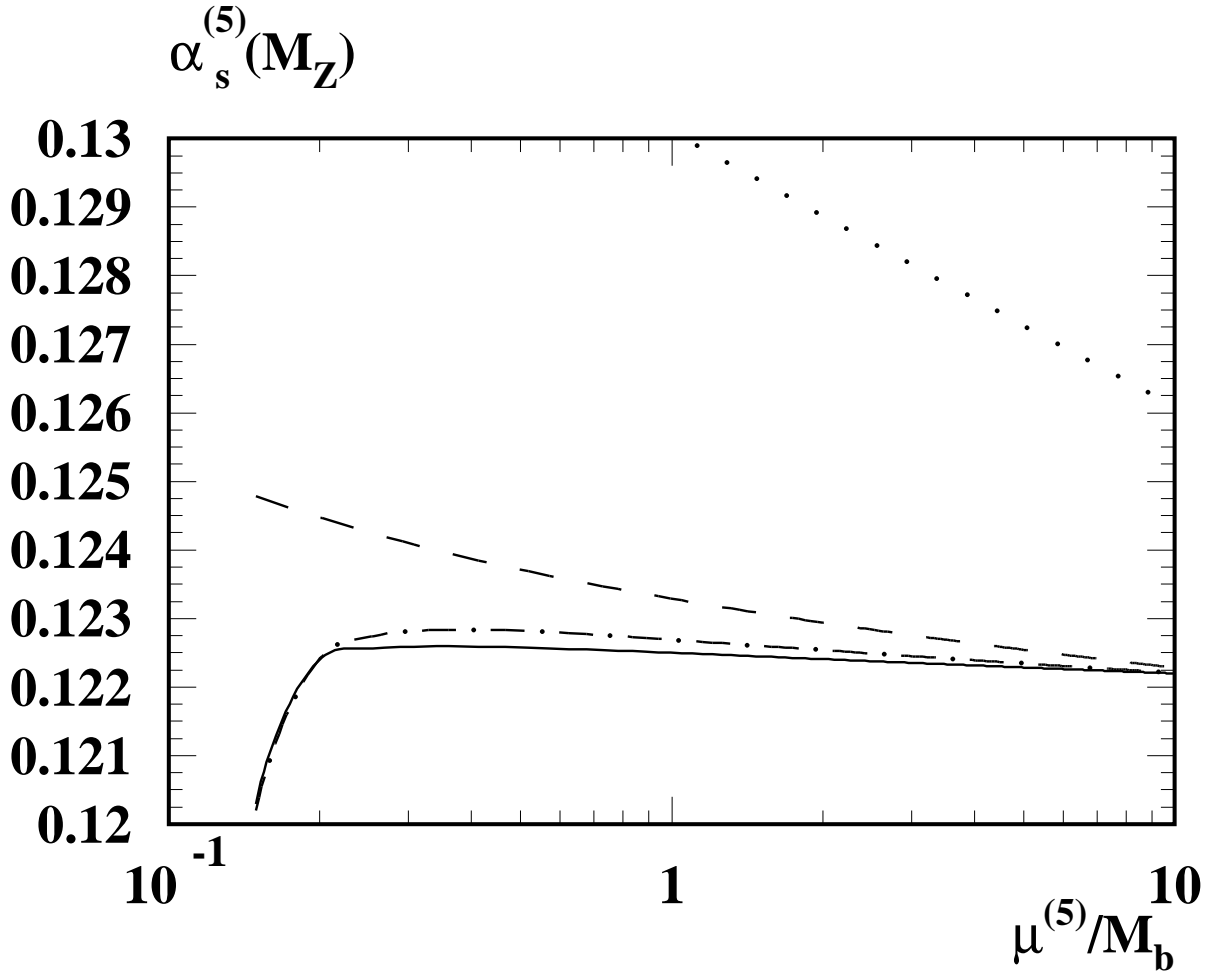}
\smallskip
\caption{$\mu^{(5)}$ dependence of $\alpha_s^{(5)}(M_Z)$ calculated from
$\alpha_s^{(4)}(M_\tau)=0.36$ and $M_b=4.7$~GeV using Eq.~(\ref{eqrgea}) at
one (dotted), two (dashed), three (dot-dashed), and four (solid) loops in
connection with Eqs.~(\ref{eqdecg}) and (\ref{eqzetagos}) at the respective
orders.}
\label{f3}
\end{figure}

Going to higher orders, one expects, on general grounds, that the relation
between $\alpha_s^{(n_l)}(\mu^\prime)$ and $\alpha_s^{(n_f)}(\mu)$, where
$\mu^\prime\ll\mu^{(n_f)}\ll\mu$, becomes insensitive to the choice of the
matching scale, $\mu^{(n_f)}$, as long as $\mu^{(n_f)}={\cal O}(m_h)$.
This has been checked in Ref.~\cite{Rod93} for three-loop evolution in
connection with two-loop matching.
Armed with our new results, we are in a position to explore the situation at 
the next order.
As an example, we consider the crossing of the bottom-quark threshold.
In particular, we wish to study how the $\mu^{(5)}$ dependence of the relation
between $\alpha_s^{(4)}(M_\tau)$ and $\alpha_s^{(5)}(M_Z)$ is reduced as we
implement four-loop evolution with three-loop matching.
Our procedure is as follows.
We first calculate $\alpha_s^{(4)}(\mu^{(5)})$ by exactly integrating
Eq.~(\ref{eqrgea}) with the initial condition $\alpha_s^{(4)}(M_\tau)=0.36$
\cite{Rod93}, then obtain $\alpha_s^{(5)}(\mu^{(5)})$ from Eqs.~(\ref{eqdecg})
and (\ref{eqzetagos}) with $M_b=4.7$~GeV, and finally compute
$\alpha_s^{(5)}(M_Z)$ with Eq.~(\ref{eqrgea}).
For consistency, $N$-loop evolution must be accompanied by $(N-1)$-loop 
matching, i.e.\ if we omit terms of ${\cal O}(\alpha_s^{N+2})$ on the 
right-hand side of Eq.~(\ref{eqrgea}), we need to discard those of
${\cal O}(\alpha_s^N)$ in Eq.~(\ref{eqzetagos}) at the same time.
In Fig.~\ref{f3}, the variation of $\alpha_s^{(5)}(M_Z)$ with $\mu^{(5)}/M_b$
is displayed for the various levels of accuracy, ranging from one-loop to
four-loop evolution.
For illustration, $\mu^{(5)}$ is varied rather extremely, by almost two orders
of magnitude.
While the leading-order result exhibits a strong logarithmic behaviour, the
analysis is gradually getting more stable as we go to higher orders.
The four-loop curve is almost flat for $\mu^{(5)}\agt1$~GeV.
Besides the $\mu^{(5)}$ dependence of $\alpha_s^{(5)}(M_Z)$, also its absolute 
normalization is significantly affected by the higher orders.
At the central matching scale $\mu^{(5)}=M_b$, we encounter a rapid, monotonic
convergence behaviour.

\begin{figure}[t]
\epsfxsize=16cm
\epsffile[99 275 463 562]{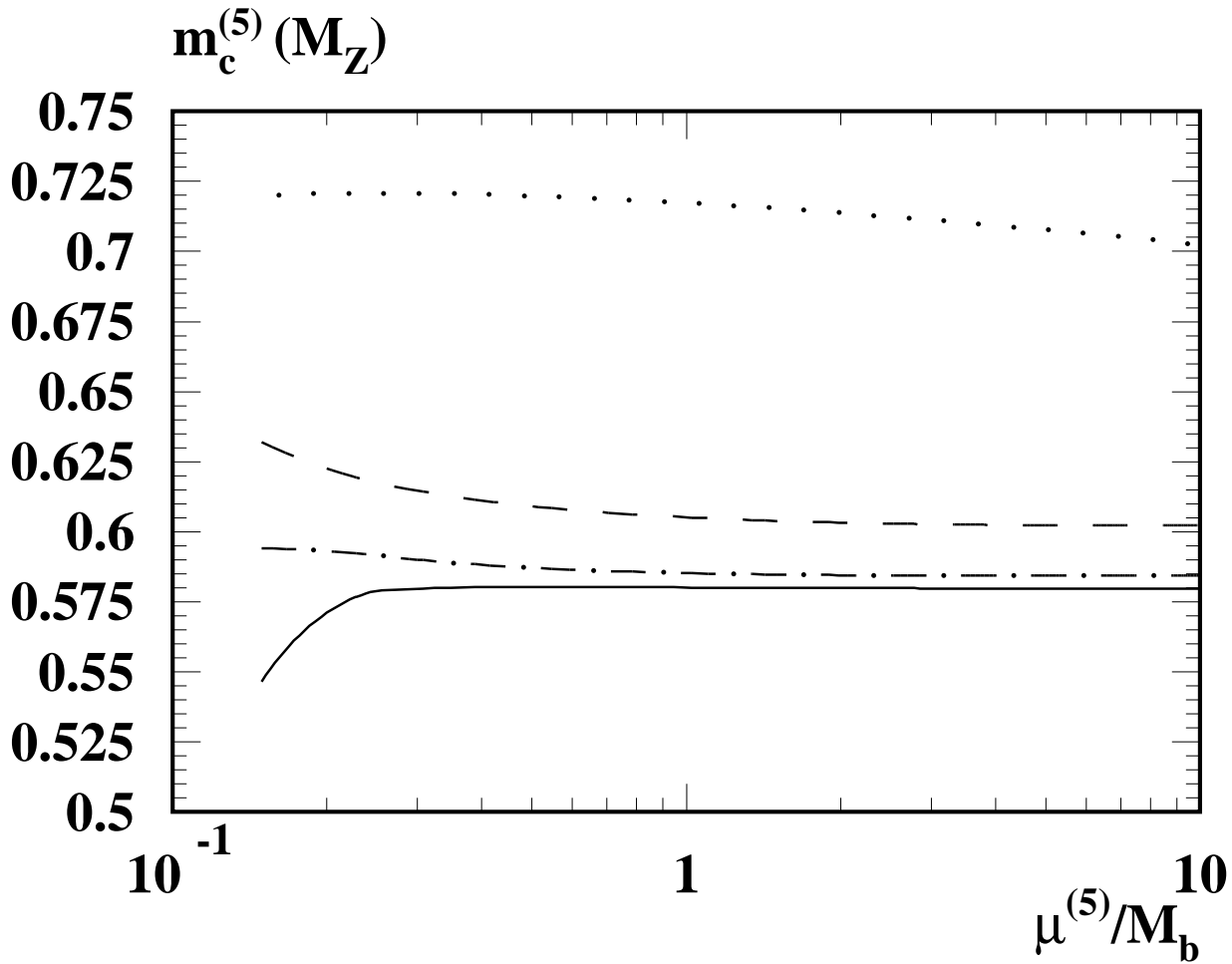}
\smallskip
\caption{$\mu^{(5)}$ dependence of $m_c^{(5)}(M_Z)$ calculated from
$\mu_c=m_c^{(4)}(\mu_c)=1.2$~GeV, $M_b=4.7$~GeV, and
$\alpha_s^{(5)}(M_Z)=0.118$ using Eq.~(\ref{eqmas}) at one (dotted), two
(dashed), three (dot-dashed), and four (solid) loops in connection with
Eqs.~(\ref{eqdecm}) and (\ref{eqzetamos}) at the respective orders.}
\label{f4}
\end{figure}

Similar analyses may be performed for the light-quark masses as well.
For illustration, let us investigate how the $\mu^{(5)}$ dependence of the
relation between $\mu_c=m_c^{(4)}(\mu_c)$ and $m_c^{(5)}(M_Z)$ changes under
the inclusion of higher orders in evolution and matching.
As typical input parameters, we choose $\mu_c=1.2$~GeV, $M_b=4.7$~GeV, and
$\alpha_s^{(5)}(M_Z)=0.118$.
We first evolve $m_c^{(4)}(\mu)$ from $\mu=\mu_c$ to $\mu=\mu^{(5)}$ via
Eq.~(\ref{eqmas}), then obtain $m_c^{(5)}(\mu^{(5)})$ from Eqs.~(\ref{eqdecm})
and (\ref{eqzetamos}), and finally evolve $m_c^{(5)}(\mu)$ from
$\mu=\mu^{(5)}$ to $\mu=M_Z$ via Eq.~(\ref{eqmas}).
In all steps, $\alpha_s^{(n_f)}(\mu)$ is evaluated with the same values of
$n_f$ and $\mu$ as $m_c^{(n_f)}(\mu)$.
In Fig.~\ref{f4}, we show the resulting values of $m_c^{(5)}(M_Z)$ 
corresponding to $N$-loop evolution with $(N-1)$-loop matching for 
$N=1,\ldots,4$.
Similarly to Fig.~\ref{f3}, we observe a rapid, monotonic convergence
behaviour at the central matching scale $\mu^{(5)}=M_b$.
Again, the prediction for $N=4$ is remarkably stable under the variation of
$\mu^{(5)}$ as long as $\mu^{(5)}\agt1$~GeV.

An interesting and perhaps even surprising aspect of the decoupling constants
$\zeta_g$ and $\zeta_m$ is that they carry the full information about the
virtual heavy-quark effects on the Higgs-boson couplings to gluons and light
quarks, respectively.
In fact, the relevant coefficient functions $C_1$ and $C_2$ in the 
renormalized version of the effective Lagrangian~(\ref{eqeff}) emerge from
$\zeta_g$ and $\zeta_m$, respectively, through logarithmic differentiation
w.r.t.\ the heavy-quark mass $m_h$.
In the LET's~(\ref{eqc1let1}) and (\ref{eqc2let}), these relationships have
been established to all orders.
By virtue of these LET's, we have succeeded in obtaining the four-loop
${\cal O}(\alpha_s^4)$ corrections to $C_1$ and $C_2$ from the three-loop
${\cal O}(\alpha_s^3)$ expressions for $\zeta_g$ and $\zeta_m$ complemented by
the four-loop results for the $\beta$ \cite{LarRitVer971} and $\gamma_m$
\cite{Che97LarRitVer972} functions.
The pure $\overline{\mbox{MS}}$ expressions for $C_1$ and $C_2$ may be found
in Eqs.~(\ref{eqc1}) and (\ref{eqc2}), respectively; the corresponding
versions written in terms of the heavy-quark pole mass $M_h$ are listed in
Eqs.~(\ref{eqc1os}) and (\ref{eqc2os}).
 
From the phenomenological point of view, $C_1$ is particularly important,
since it enters the theoretical prediction for the cross section of the
$gg\to H$ subprocess, which is expected to be the dominant production 
mechanism of the SM Higgs boson $H$ at the CERN proton-proton collider LHC.
The ${\cal O}(\alpha_s^2)$ term of Eq.~(\ref{eqc1}) is routinely included in
next-to-leading-order calculations within the parton model \cite{Daw91}.
Recently, a first step towards the next-to-next-to-leading order has been
taken in Refs.~\cite{Kra96,Spi97} by considering the resummation of
soft-gluon radiation in $pp\to H+X$.
The ${\cal O}(\alpha_s^3)$ formula~(13) for $\kappa_{h,H}$ in the revised
version of Ref.~\cite{Kra96} agrees with our Eq.~(\ref{eqc1}) for
$C_1^{(n_f)}(\mu)$ to this order, if we identify
\begin{eqnarray}
{\cal C}_1(\mu_t,M_H)&\n=\n&
\frac{\alpha_s^{(5)}(\mu_t)\,\beta^{(5)}\left(\alpha_s^{(5)}(M_H)\right)}
{\alpha_s^{(5)}(M_H)\,\beta^{(5)}\left(\alpha_s^{(5)}(\mu_t)\right)}
C_1^{(6)}(\mu_t)
\nonumber\\
&\n=\n&
-\frac{1}{12}\,\frac{\alpha_s^{(5)}(M_H)}{\pi}\sqrt{\kappa_{h,H}}.
\label{eqkap}
\end{eqnarray}
The RG improvement of $C_1$ in the first line of Eq.~(\ref{eqkap}) is adopted
from Eq.~(30) of Ref.~\cite{CheKniSte96}, where it has been employed in the
context of three-loop ${\cal O}(\alpha_s^2G_FM_t^2)$ corrections to hadronic
Higgs-boson decays.
We stress that the particular choice of scales in Eq.~(\ref{eqkap}) is
essential in order to recover Eq.~(13) of Ref.~\cite{Kra96} from the general
framework elaborated here.
The $\mu$ dependence of $\kappa_{h,H}$ is not displayed in Ref.~\cite{Kra96}.
We note that the ${\cal O}(\alpha_s^3)$ result for $C_1$ was originally
obtained in Ref.~\cite{CheKniSte97}.

In their Eq.~(8), the authors of Ref.~\cite{Kra96} also present an alternative
version of a generic formula for $C_1$; see also Eq.~(25) of Ref.~\cite{Spi97}.
A central ingredient of this formula is the quantity $\beta_t(\alpha_s)$,
which they paraphrase as the top-quark contribution to the QCD $\beta$
function at vanishing momentum transfer.
According to Ref.~\cite{SpiPC}, $\beta_t(\alpha_s)$ may be obtained from some
gauge-independent variant of the top-quark contribution to the bare gluon
self-energy at vanishing momentum transfer.
Since we are unable to locate in the literature, including
Refs.~\cite{Kra96,Spi97} and the papers cited therein, a proper general
definition of $\beta_t(\alpha_s)$ in terms of the familiar quantities of QCD,
it remains unclear whether Eq.~(8) of Ref.~\cite{Kra96} agrees with our
Eq.~(\ref{eqc1let}) in higher orders.
If we were to fix $\beta_t(\alpha_s)$ so that Eq.~(8) of Ref.~\cite{Kra96}
reproduces our Eq.~(\ref{eqc1let}) after the RG-improvement of
Eq.~(\ref{eqkap}), the result would be
\begin{equation}
\beta_t\left(\mu,\alpha_s^{(6)},m_t^{(6)}\right)
=2\pi\left[\beta^{(5)}\left(\alpha_s^{(5)}\right)
-\beta^{(6)}\left(\alpha_s^{(6)}\right)
\frac{\partial\alpha_s^{(5)}}{\partial\alpha_s^{(6)}}\right],
\end{equation}
where
$\alpha_s^{(5)}=\alpha_s^{(5)}\left(\mu,\alpha_s^{(6)},m_t^{(6)}\right)$.
We emphasize that, in order to derive Eq.~(\ref{eqc1let}), we found it 
indispensable to consider the effective Lagrangian~(\ref{eqeff}) comprising a
complete basis of scalar dimension-four operators.
As a result, Eq.~(\ref{eqc1let}) relates $C_1$ to basic, gauge-invariant
quantities of QCD in a simple way, and is manifestly valid to all orders.

The formalism developed for QCD in Sections~\ref{seccon} and \ref{seclet}
naturally carries over to QED.
In particular, it allows us to evaluate the two-loop
${\cal O}(\bar\alpha\alpha_s)$ and three-loop ${\cal O}(\bar\alpha\alpha_s^2)$
corrections to the decoupling constant $\zeta_{g\gamma}$ for the
fine-structure constant $\bar\alpha(\mu)$ renormalized according to the
$\overline{\mbox{MS}}$ scheme.
The results formulated in terms of $m_h(\mu)$ and $M_h$ may be found in
Eqs.~(\ref{eqzetagg}) and (\ref{eqzetaggos}), respectively.
An appropriately modified LET relates the coefficient function $C_{1\gamma}$
of the heavy-quark effective $\gamma\gamma H$ coupling to $\zeta_{g\gamma}$.
Similarly to the case of the $ggH$ coupling, this enables us to calculate
$C_{1\gamma}$ through four loops, i.e.\ to ${\cal O}(\bar\alpha\alpha_s^n)$
with $n=0,\ldots,3$.
The final results written with $m_h(\mu)$ and $M_h$ are listed in
Eqs.~(\ref{eqc1gamma}) and (\ref{eqc1gammaos}), respectively.

\vspace{1cm}
\noindent
{\bf Note added}
\smallskip

After the submission of this manuscript, a preprint \cite{Rod97} has appeared
in which $\alpha_s^{(5)}(M_Z)$ is related to $\alpha_s^{(3)}(M_\tau)$ with
four-loop evolution \cite{LarRitVer971} and three-loop matching
\cite{CheKniSte97als} at the charm- and bottom-quark thresholds.
In particular, the theoretical error due to the evolution procedure is
carefully estimated.
In Ref.~\cite{Rod97}, also the logarithmic terms of the three-loop decoupling
constant $\zeta_g$, which was originally found in Ref.~\cite{CheKniSte97als},
are confirmed using standard RG techniques.
These techniques were also employed in Ref.~\cite{CheKniSte97als} in order
to check the logarithmic part of the full result, which was obtained there
through explicit diagrammatical calculation.

\vspace{1cm}
\noindent
{\bf Acknowledgements}
\smallskip

We thank Werner Bernreuther for a clarifying communication regarding 
Ref.~\cite{BerWet82} and Michael Spira for carefully reading this manuscript
and for his comments.
The work of K.G.C. was supported in part by INTAS under Contract
INTAS--93--744--ext.

\newpage
\noindent
{\Large\bf Appendix}

\renewcommand {\theequation}{\Alph{section}.\arabic{equation}}
\begin{appendix}

\setcounter{equation}{0}
\boldmath
\section{\label{appnc}Results for the gauge group SU($N_c$)}
\unboldmath

In the following, we list the decoupling constants $\zeta_m$ and $\zeta_g$
appropriate for the general gauge group SU($N_c$).
The results read 
\begin{eqnarray}
\zeta_m &\n=\n&1
+\left(\frac{\alpha_s^{(n_f)}(\mu)}{\pi}\right)^2\left(\frac{1}{N_c}-N_c\right)
\left(-\frac{89}{1152}+\frac{5}{96}\ln\frac{\mu^2}{m_h^2}
-\frac{1}{32}\ln^2\frac{\mu^2}{m_h^2}\right)
\nonumber\\
&\n\n&{}
+\left(\frac{\alpha_s^{(n_f)}(\mu)}{\pi}\right)^3
\Bigg\{
\frac{1}{N_c^2}\left(
-\frac{683}{4608}
+\frac{57}{256}\zeta(3)
-\frac{9}{64}\zeta(4)
+\frac{1}{32}B_4\right)
\nonumber\\
&\n\n&{}
+\frac{1}{N_c}\left(
\frac{1685}{62208}
-\frac{7}{144}\zeta(3)\right)
+\frac{907}{31104}
-\frac{397}{2304}\zeta(3)
-\frac{1}{32}B_4
\nonumber\\
&\n\n&{}
+N_c\left(
-\frac{1685}{62208}
+\frac{7}{144}\zeta(3)\right)
+N_c^2\left(
\frac{14813}{124416}
-\frac{29}{576}\zeta(3)
+\frac{9}{64}\zeta(4)\right)
\nonumber\\
&\n\n&{}
+\left[\frac{1}{N_c^2}\left(
-\frac{13}{512}
+\frac{3}{32}\zeta(3)\right)
+\frac{31}{864N_c}
+\frac{1}{32}
-\frac{31}{864}N_c
-N_c^2\left(
\frac{3}{512}
+\frac{3}{32}\zeta(3)\right)
\right]\ln\frac{\mu^2}{m_h^2} 
\nonumber\\
&\n\n&{}
+\left(-\frac{1}{32N_c^2}
-\frac{5}{576N_c}
-\frac{5}{384}
+\frac{5}{576}N_c
+\frac{17}{384}N_c^2\right)
\ln^2\frac{\mu^2}{m_h^2}
\nonumber\\
&\n\n&{}
+\left(\frac{1}{144N_c}
-\frac{11}{576}
-\frac{1}{144}N_c
+\frac{11}{576}N_c^2\right) 
\ln^3\frac{\mu^2}{m_h^2}
\nonumber\\
&\n\n&{}
+n_l\left(\frac{1}{N_c}-N_c\right)\left(
-\frac{1327}{31104}
+\frac{1}{36}\zeta(3)
+\frac{53}{1152}\ln\frac{\mu^2}{m_h^2}
+\frac{1}{288}\ln^3\frac{\mu^2}{m_h^2}\right)
\Bigg\},
\nonumber\\
\zeta_g^2&\n=\n&1
+\frac{\alpha_s^{(n_f)}(\mu)}{\pi}
\left(-\frac{1}{6}\ln\frac{\mu^2}{m_h^2}\right)
\nonumber\\
&\n\n&{}
+\left(\frac{\alpha_s^{(n_f)}(\mu)}{\pi}\right)^2
\left[
\frac{13}{192N_c}
+\frac{25}{576}N_c
-\left(\frac{1}{16N_c}
+\frac{7}{48}N_c\right)
\ln\frac{\mu^2}{m_h^2}
+\frac{1}{36}\ln^2\frac{\mu^2}{m_h^2}\right]
\nonumber\\
&\n\n&{}
+\left(\frac{\alpha_s^{(n_f)}(\mu)}{\pi}\right)^3
\Bigg\{
\frac{1}{N_c^2}\left(
-\frac{97}{2304}
+\frac{95}{1536}\zeta(3)\right)
+\frac{1}{N_c}\left(
-\frac{103}{10368}
+\frac{7}{512}\zeta(3)\right)
\nonumber\\
&\n\n&{}
-\frac{1063}{5184}
+\frac{893}{3072}\zeta(3)
+N_c\left(
\frac{451}{20736}
-\frac{7}{256}\zeta(3)\right)
+N_c^2\left(
\frac{7199}{13824}
-\frac{17}{48}\zeta(3)
\right)
\nonumber\\
&\n\n&{}
+\left(-\frac{9}{256N_c^2}
-\frac{5}{192N_c}
-\frac{119}{1152}
-\frac{23}{3456}N_c
-\frac{1169}{6912}N_c^2\right)
\ln\frac{\mu^2}{m_h^2}
\nonumber\\
&\n\n&{}
+\left(\frac{5}{192N_c}
-\frac{11}{384}
+\frac{35}{576}N_c
-\frac{1}{128}N_c^2\right)
\ln^2\frac{\mu^2}{m_h^2}
-\frac{1}{216}\ln^3\frac{\mu^2}{m_h^2}
\nonumber\\
&\n\n&{}
+n_l\Bigg[
\frac{41}{1296N_c}
-\frac{329}{10368}N_c
+\left(-\frac{5}{384N_c}
+\frac{139}{3456}N_c\right)
\ln\frac{\mu^2}{m_h^2}
\nonumber\\
&\n\n&{}
+\left(\frac{1}{96N_c}
-\frac{1}{96}N_c\right)
\ln^2\frac{\mu^2}{m_h^2}\Bigg]\Bigg\}.
\end{eqnarray}
For $N_c=3$, we recover Eqs.~(\ref{eqzetam}) and (\ref{eqzetag}).

\setcounter{equation}{0}
\section{\label{appz2z3}Decoupling relations for the quark and gluon fields}

In analogy to Eqs.~(\ref{eqdecg}) and (\ref{eqdecm}), the renormalized
decoupling constants $\zeta_2$ and $\zeta_3$ for the quark and gluon fields, 
respectively, arise from the relations
\begin{eqnarray}
\psi_q^\prime 
&\n=\n& 
\sqrt{\frac{Z_2}{Z_2^\prime}\zeta_2^0}\,\psi_q
=\sqrt{\zeta_2}\,\psi_q,
\nonumber\\
G_\mu^{\prime a} &\n=\n& 
\sqrt{\frac{Z_3}{Z_3^\prime}\zeta_3^0}\,G_\mu^a
=\sqrt{\zeta_3}\,G_\mu^a.
\end{eqnarray}
Of course, $\zeta_2$ and $\zeta_3$ are both gauge dependent.
Restricting ourselves to the case $N_c=3$, we find in the covariant
gauge~(\ref{eqcov})
\begin{eqnarray}
\zeta_2&\n=\n&1
+\left(\frac{\alpha_s^{(n_f)}(\mu)}{\pi}\right)^2
\Bigg({5\over 144}
-{1\over 12}\ln\frac{\mu^2}{m_h^2}\Bigg) 
+\left(\frac{\alpha_s^{(n_f)}(\mu)}{\pi}\right)^3\Bigg[
{42811\over 62208}
+{1\over 18}\zeta(3)
\nonumber\\
&\n\n&{}
-{155\over 192}\ln\frac{\mu^2}{m_h^2}
+{49\over 576}\ln^2\frac{\mu^2}{m_h^2}
-{1\over 96}\ln^3\frac{\mu^2}{m_h^2}
+n_l\Bigg(
{35\over 3888}
+{5\over 432}\ln\frac{\mu^2}{m_h^2}\Bigg) 
\nonumber\\
&\n\n&{}
+\xi\Bigg(
-{2387\over 6912}
+{1\over 12}\zeta(3)
+{121\over 576}\ln\frac{\mu^2}{m_h^2}
-{13\over 192}\ln^2\frac{\mu^2}{m_h^2}
+{1\over 96}\ln^3\frac{\mu^2}{m_h^2}\Bigg)\Bigg],
\nonumber\\
\zeta_3&\n=\n&1
+\frac{\alpha_s^{(n_f)}(\mu)}{\pi}\Bigg(
{1\over 6}\ln\frac{\mu^2}{m_h^2}\Bigg) 
+ \left(\frac{\alpha_s^{(n_f)}(\mu)}{\pi}\right)^2\Bigg(
{91\over 1152}
+{29\over 96}\ln\frac{\mu^2}{m_h^2}
+{3\over 32}\ln^{2}\frac{\mu^2}{m_h^2}\Bigg) 
\nonumber\\
&\n\n&{}
+\left(\frac{\alpha_s^{(n_f)}(\mu)}{\pi}\right)^3\Bigg[
-{284023\over 62208}
+{86183\over 27648}\zeta(3)
+{99\over 128}\zeta(4)
-{1\over 32}B_4
\nonumber\\
&\n\n&{}
+\left({52433\over 27648}
-{33\over 64}\zeta(3)\right)\ln\frac{\mu^2}{m_h^2}
+{383\over 2304}\ln^2\frac{\mu^2}{m_h^2}
+{119\over 768}\ln^3\frac{\mu^2}{m_h^2}
\nonumber\\
&\n\n&{}
+n_l\Bigg(
{3307\over 15552}
-{1\over 12}\zeta(3)
-{293\over 1152}\ln\frac{\mu^2}{m_h^2}
+{1\over 36}\ln^2\frac{\mu^2}{m_h^2}
-{1\over 96}\ln^3\frac{\mu^2}{m_h^2}\Bigg) 
\nonumber\\
&\n\n&{}
+\xi\Bigg(
-{677\over 1536}
+{3\over 32}\zeta(3)
+{233\over 1024}\ln\frac{\mu^2}{m_h^2}
-{3\over 32}\ln^2\frac{\mu^2}{m_h^2}
+{3\over 256}\ln^3\frac{\mu^2}{m_h^2}\Bigg)\Bigg].
\end{eqnarray}

\end{appendix}

\end{document}